\documentclass{ar-1col}

\usepackage{url, natbib, verbatim, ulem}
\usepackage{amsmath,amssymb}
\usepackage[flushleft]{threeparttable}

\jname{Annual Review of Astronomy and Astrophysics}
\jvol{AA}
\jyear{2020}
\doi{10.1146/((please add article doi))}

\newcommand{\lya}{Ly$\alpha$}

\newcommand{\msun}{\ensuremath{{\rm M}_\odot}}

\newcommand{\laf}{\lya\ forest}

\newcommand\ion[2]{#1$\;${\scshape{#2}}}

\newcommand{\HI}{\ensuremath{\mbox{\ion{H}{i}}}}
\newcommand{\Hatom}{\ensuremath{{\rm H}^0}}
\newcommand{\Hmol}{\ensuremath{{\rm H}_2}}
\newcommand{\xHI}{\ensuremath{x({\rm H}^0)}}
\newcommand{\HII}{\ensuremath{\mbox{\ion{H}{ii}}}}

\newcommand{\NHI}{\ensuremath{N(\mbox{\ion{H}{i}})}}
\newcommand{\logNHI}{\ensuremath{\log N(\mbox{\ion{H}{i}})}}
\newcommand{\MtoH}{\ensuremath{[{\rm M/H}]}}
\newcommand{\ICF}{\ensuremath{{\rm ICF}}}

\newcommand{\OmegaGas}{\ensuremath{\Omega_{\rm neutral \ gas}}}
\newcommand{\OmegaMet}{\ensuremath{\Omega_{\rm metals}}}
\newcommand{\OmegaDust}{\ensuremath{\Omega_{\rm dust}}}
\newcommand{\OmegaMol}{\ensuremath{\Omega_{\rm molecular \ gas}}}
\newcommand{\OmegaBaryons}{\ensuremath{\Omega_{\rm baryons}}}
\newcommand{\rhocrit}{\ensuremath{\rho_{\rm crit, 0}}}
\newcommand{\rhostar}{\ensuremath{\rho_\bigstar}}
\newcommand{\rhomet}{\ensuremath{\rho_{\rm metals}}}
\newcommand{\rhogas}{\ensuremath{\rho_{\rm neutral \ gas}}}

\newcommand{\dtg}{\ensuremath{{\rm DTG}}}
\newcommand{\dtm}{\ensuremath{{\rm DTM}}}
\newcommand{\deltaX}{\ensuremath{\delta_X}}
\newcommand{\deltaFe}{\ensuremath{\delta_{\rm Fe}}}
\newcommand{\Tdep}{\ensuremath{\tau_{\rm dep}}}

\newcommand{\la}{\lesssim}
\newcommand{\ga}{\gtrsim}
\newcommand{\column}{\ensuremath{{\rm cm}^{2}}}

\def\lya{Ly-$\alpha$}

\def\hi{H~{\sc i}}
\def\hii{H~{\sc ii}}

\def\lognhi{\ensuremath{\log N(\mbox{\ion{H}{i}})}}

\def\civ{C~{\sc iv}}

\def\mgii{Mg~{\sc ii}}

\def\siiv{Si~{\sc iv}}

\def\oi{O~{\sc i}}
\def\ovi{O~{\sc vi}}

\def\nv{N~{\sc v}}

\begin{document}

\markboth{P\'eroux \& Howk}{The Cosmic Baryon and Metal Cycles}

\title{The Cosmic Baryon and Metal Cycles}

\author{C\'eline P\'eroux$^{1, 2}$ and J. Christopher Howk$^3$
\affil{$^1$European Southern Observatory, Karl-Schwarzschild-Str. 2, 85748 Garching-bei-M\"unchen, Germany; email: cperoux@eso.org }
\affil{$^2$Aix Marseille Universit\'e, CNRS, LAM (Laboratoire d'Astrophysique de Marseille) UMR 7326, 13388, Marseille, France}
\affil{$^3$Department of Physics, University of Notre Dame, Notre Dame, IN 46556, USA; email: jhowk@nd.edu}}

\begin{abstract}

Characterizing the relationship between stars, gas, and metals in galaxies is a critical component of understanding the cosmic baryon cycle. We compile contemporary censuses of the baryons in collapsed structures, their chemical make-up and dust content. We show that:

\noindent $\bullet$ The \HI\ mass density of the Universe is well determined to redshifts $z\approx5$ and shows minor evolution with time. New observations of molecular hydrogen reveal its evolution mirrors that of the global star formation rate density. The constant cosmic molecular gas depletion timescale points to a universal relationship between gas reservoirs and star formation.

\noindent $\bullet$ The metal mass density in cold gas ($T \la 10^4$ K) contains virtually all the metals produced by stars for $z \ga 2.5$. At lower redshifts, the contributors to the total amount of metals are more diverse; at $z<1$, most of the observed metals are bound in stars. Overall there is little evidence for a ``missing metals problem'' in modern censuses.

\noindent $\bullet$ We characterize the dust content of neutral gas over cosmic time, finding the dust-to-gas and dust-to-metals ratios fall with decreasing metallicity. We calculate the cosmological dust mass density in the neutral gas up to $z\approx5$. There is good agreement between multiple tracers of the dust content of the Universe.

\end{abstract}

\begin{keywords}
baryon density,
atomic and molecular gas,
cosmic abundances,
galaxy chemical evolution,
interstellar dust,
quasar absorption line spectroscopy
%
\end{keywords}
\maketitle

\tableofcontents

\section{CONTEXT}
\label{sec:context}

\subsection{Motivation of this Review}
\label{sec:motivation}

Astronomers now know the basic constituents of the present Universe: 73\% dark
energy, 23\% dark matter, 4\% in baryons. The term {\it baryons} is used to refer to the normal matter of the Universe (inclusive of the baryonic
and leptonic matter discussed by physicists). The baryon density parameter --
the comoving baryon density normalized to the critical density, $\OmegaBaryons
\equiv \rho_{\rm baryons} / \rhocrit$ -- that describes the quantity of normal
matter is a fundamental cosmological parameter. One of the great successes of
the last decades is the excellent agreement between estimates of \OmegaBaryons\
from primordial nucleosynthesis \citep{cooke2018} and Cosmic Microwave
Background anisotropies \citep{planck2016}. From the Big Bang onwards, the
baryons collapse with dark matter to form the large-scale structure, galaxies,
and stars we observe.
While their mass grows with time, only a minority of the baryonic matter is
found in stars -- even today, some 13.7 Gyr after the Big Bang, $>90\%$ of
the baryons are found in the gaseous phase of the Universe. The gas, notably the
cold gas traced by \HI\ and \Hmol, provides the reservoir of fuel for forming
stars.

\begin{marginnote}[]
    \entry{Baryons}{Normal matter}
    \entry{Metals}{Elements heavier than helium}
    \entry{Metallicity}{Relative quantity of metal atoms compared to hydrogen}
    \entry{Dust}{Solid-phase materials composed of metals}
\end{marginnote}

The formation of stars leads naturally to the creation of {\it metals} -- the
elements heavier than helium. The study of the metal content in cosmic
environments provides fundamental insights into the evolutionary processes that
drive the formation and evolution of galaxies and the stars within them.  Metals
are largely created in stellar interiors through nuclear fusion reactions. They
are expelled from stars through the supernova explosions that mark the ends of
massive stars' lives. The collective effects of supernovae also serve to distribute the metals on much
larger scales: mixing metals into the interstellar medium (ISM) of galaxies and
even beyond. The metals that remain in the ISM are incorporated into the next
generation of stars. This includes the metals chemically sequestered into dust
grains, the solid phase material that plays a central role in catalyzing
formation of the molecules essential to star formation.

Observations of the temporal and spatial evolution of metals are key to
constraining this global picture of the Universe's evolution. Early accounting
of the total metal budget found an order of magnitude shortfall in the comoving
density of observed metals compared with those expected to be produced by the
stellar content of the Universe \citep{pettini1999, pagel1999}.  This so-called
{\it ``missing metals problem''} was originally based on metal abundances
derived for components of the Universe at $z=2.5$, including gas traced by
quasar absorbers and galaxies. It was then postulated that the missing metals
would be found far from their production sites, in the hot gas filling galactic
halos and proto-clusters.  This problem has been revisited several times
\citep[e.g.,][]{ferrara2005, bouche2005, bouche2006, bouche2007, shull2014}.

\begin{marginnote}[]
    \entry{ISM}{Interstellar Medium}
    \entry{CGM}{Circumgalactic Medium}
    \entry{IGM}{Intergalactic Medium}
    \entry{WHIM}{Warm-Hot Intergalactic Medium}
\end{marginnote}

This review synthesizes modern censuses of gas, metals, and dust in the baryonic
components most closely related to galaxies and their star formation. By
gathering the latest robust observations and state-of-the-art, physically
motivated cosmological hydrodynamical models of galaxy formation, we assess
these quantities continuously with look-back time over 90\% of the age of the
Universe. Given the limited space available, it is impossible to provide a thorough survey
of such a huge community effort without leaving out significant contributions or
whole subfields. Specifically in this volume, \cite{tacconi2020} offers an alternative view of
the molecular and dust content of galaxies, and \cite{forsterschreiber2020}
focuses on the kinematics and resolved metal content of intermediate redshift
galaxies. In what follows we focus on a series of questions: How does the gas
reservoir which fuels star formation evolve (\S\ref{sec:baryon})? Where are the
metals in the Universe (\S\ref{sec:metal})? What is the cosmic evolution of
dust mass (\S\ref{sec:dust})? In the final section, \S\ref{sec:future}, we
provide a forward looking view of the field.

All results presented here assume a concordance $\Lambda$CDM cosmology with
parameters ($\Omega_M$, $\Omega_{\Lambda}$, $h$) = (0.3, 0.7, 0.7), where
$\Omega_M$ and $\Omega_{\Lambda}$ are the normalized mass-energy densities of
matter and the cosmological constant, respectively, at $z=0$ relative to the
critical density, $\rhocrit \equiv  8 \pi G/3H_0^2$. The quantity $h \equiv
H_0/({\rm 100 \ km \ s^{-1} \ Mpc^{-1}})$ is a dimensionless statement of the
present-day Hubble constant, $H_0$. We use comoving coordinates when discussing
volume densities throughout this work, such that a comoving Mpc, cMpc, is
related to the physical (proper) distance by ${\rm cMpc \equiv Mpc} \times (1+z)$. All
logarithms assume base-10.

\begin{marginnote}[]
 \entry{Mass densities}{$\rho_X$ is the {\it comoving} mass density for a component $X$ (in $\msun \ {\rm cMpc^{-3}}$).}
 \entry{Density parameters}{\\$\Omega_X$ is the density normalized by the critical density at $z=0$, $\Omega_X \equiv \rho_X/\rhocrit$ (unitless).}
\end{marginnote}

\subsection{Probing Baryons and Metals}
\label{sec:abs}

\begin{figure}[h]
\includegraphics[width=6.5in]{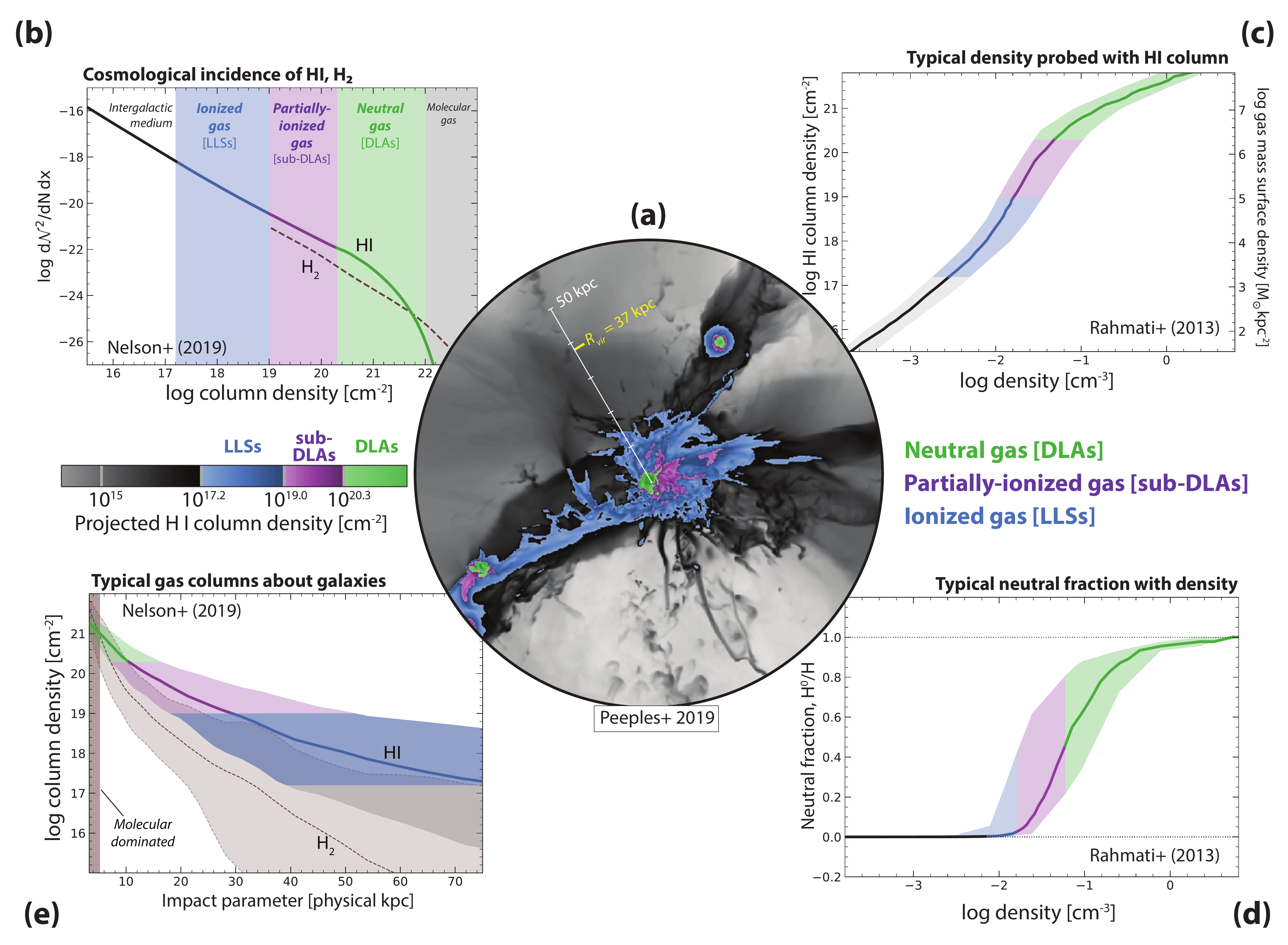}
\caption{A broad view of the physical properties of cold gas discussed in this
review and its relationship to galaxies. This figure summarizes the behavior of
gas column (surface) densities in $z \sim 2-3$ simulated galaxy halos. The
central panel (a) shows projected gas column densities for a halo that will
develop into a Milky Way-mass system by $z=0$. The \HI\ column densities are
coded according to the scale in the middle-left, and the color-coding for all
panels is summarized in the middle-right. The four sub-panels show (b) the
cosmological column density distribution, (c) the typical projected \HI\ column
density as a function of total particle density, (d) the neutral fraction of the
gas as a function of particle density, and (e) the typical column density of
\HI\ and \Hmol\ as a function of projected distance from simulated galaxies. The
rarer high column density (and thus highest density / least ionized) gas is
preferentially found in the inner regions of galaxies.  } \label{fig:sketch}
\end{figure}

Only a minority of the baryons are scrutinized by observations of starlight from
galaxies. The remaining baryons reside in low-density gas which is challenging
to detect in emission \citep{wijers2019, augustin2019}. The baryons and metals
associated with the majority of the gas that lies beyond the regions excited by
hot stars is studied using resonant absorption lines in the spectra of
background sources. This is a particularly powerful technique when using bright
background sources such as quasars, gamma ray bursts (GRBs) \citep{bolmer2019},
fast radio burst (FRBs) \citep{prochaska2019}, or even other galaxies
\citep{cooke2015, peroux2018}. Absorption lines provide a measure of the surface
density or {\it column density} of atoms (molecules) between the observer and
the background source (expressed in atoms cm$^{-2}$). The \HI\ column density,
for example, along a sightline passing through gas with a neutral hydrogen
particle density, $n_{\rm H\, I}$ (in atoms cm$^{-3}$) is
\begin{equation}
    \NHI \equiv \int n_{\rm H\, I} \, ds,
\end{equation}
where $s$ is the path the sightline takes through the gas. The important 
advantage of
absorption over emission from galaxies is the ability to reach low gas
densities. Moreover, the sensitivity of this technique is redshift independent and offers
a powerful tool to study the cosmic evolution of the baryon and metal content of
the Universe. While individual absorption measurements are limited to a
pencil-beam along the line-of-sight, the large samples now assembled allow us to
statistically measure the mean properties of galaxies by combining many lines of
sight, thereby also minimizing cosmic variance effects.

In Figure \ref{fig:sketch} we highlight the physical properties of the gas that
is studied in absorption, focusing on the high column density gas in and around
galaxies. We focus this review on these densest environments, which play a key
role in the baryon cycle of galaxies and harbor the majority of the metals in
the Universe. The central panel, Figure~\ref{fig:sketch}a, shows the projected
\HI\ column density, \NHI, around a typical simulated galaxy at $z=2$
\citep[from the Enzo-based FOGGIE suite of][]{peeples2019}.  The remaining
panels in Figure~\ref{fig:sketch} connect the gas' physical properties (notably
its density and ionization conditions) with its environment and measurable
properties (in this case drawn from the analysis of the EAGLE and IllustrisTNG
simulations in \citealt{rahmati2013} and \citealt{nelson2019}, respectively).
Clear trends emerge when examining in Figure \ref{fig:sketch}. The higher column
densities are found closer to the centers of simulated galaxies
(Figure~\ref{fig:sketch}e), and high column density gas is denser
(Figure~\ref{fig:sketch}c) and more neutral (Figure \ref{fig:sketch}d). Because
higher column density gas has a smaller sky cross section, the frequency with
which random sight lines in ``blind'' surveys intercept this gas is smaller than
for lower column densities (Figure~\ref{fig:sketch}b).

\begin{marginnote}[]
    \entry{molecular gas}{$N({\rm H}_2) \ge \NHI$}
    \entry{neutral gas (DLAs)}{\\$\NHI \ge 10^{20.3} \ {\rm cm^{-2}}$}
    \entry{partially-ionized gas (sub-DLAs)}{\\$\NHI = 10^{19.0}-10^{20.3} \ {\rm cm^{-2}}$   }
    \entry{ionized gas (LLS)}{\\$\NHI = 10^{17.2}-10^{19.0}  \ {\rm cm^{-2}}$}
\end{marginnote}

The highest \HI\ column densities in Figure~\ref{fig:sketch} trace dense gas and
are largely self-shielded from ionizing ultraviolet background photons, thus
remaining mostly neutral. We identify {\it neutral gas} regions in this review
as those having \HI\ column densities $\NHI > 2 \times
10^{20}$ cm$^{-2}$, often referred to as Damped Lyman-$\alpha$ Systems or DLAs
\citep[see review by][]{wolfe2005}. These systems trace both the cold and warm
neutral medium.\footnote{Here we are using the nomenclature ``neutral gas'' to
refer specifically to the atomic component; we consider the molecular component
of the coldest gas separately.} This canonical definition was introduced by
\citet{wolfe1986} because such column densities are characteristic of \HI\ 21-cm
measurements of local disk galaxies while providing lines of sufficient
equivalent width (\S \ref{sec:basic_abs}) to be easily detectable in the
low-resolution spectra available at the time.

Lower \HI\ surface density gas found further from the cores of galaxies (Figure
\ref{fig:sketch}e) is increasingly more ionized and multi-phase (Figure
\ref{fig:sketch}e). The ionization state of hydrogen is set by the balance of
the reaction
\begin{equation}
    H^0 + \gamma \rightleftharpoons H^+ + e^-.
\end{equation}
The optical depth at the Lyman limit ($h\nu = 1\ {\rm Ryd} \approx 13.6 \ {\rm
eV}$ or $\lambda \approx 912$ \AA) is key: high optical depths at energies $\sim
1-2$ Ryd keep ionizing photons ($\gamma$) from propagating throughout the gas.
For $10^{19} \le \NHI \le 2\times10^{20}$  cm$^{-2}$, the gas contains a mixture
of neutral and ionized H, and we refer to such regions as {\it partially-ionized
gas} throughout this review. This column density regime corresponds to the {\it
sub-DLAs} first denoted by \citet{peroux2003a}, who set the column density
definition to include all systems which contribute significantly to the total
\HI\ mass density. These systems are sometimes referred to as {\it super Lyman-limit
systems}.

\begin{figure}[h]
\includegraphics[width=6.in, angle=0]{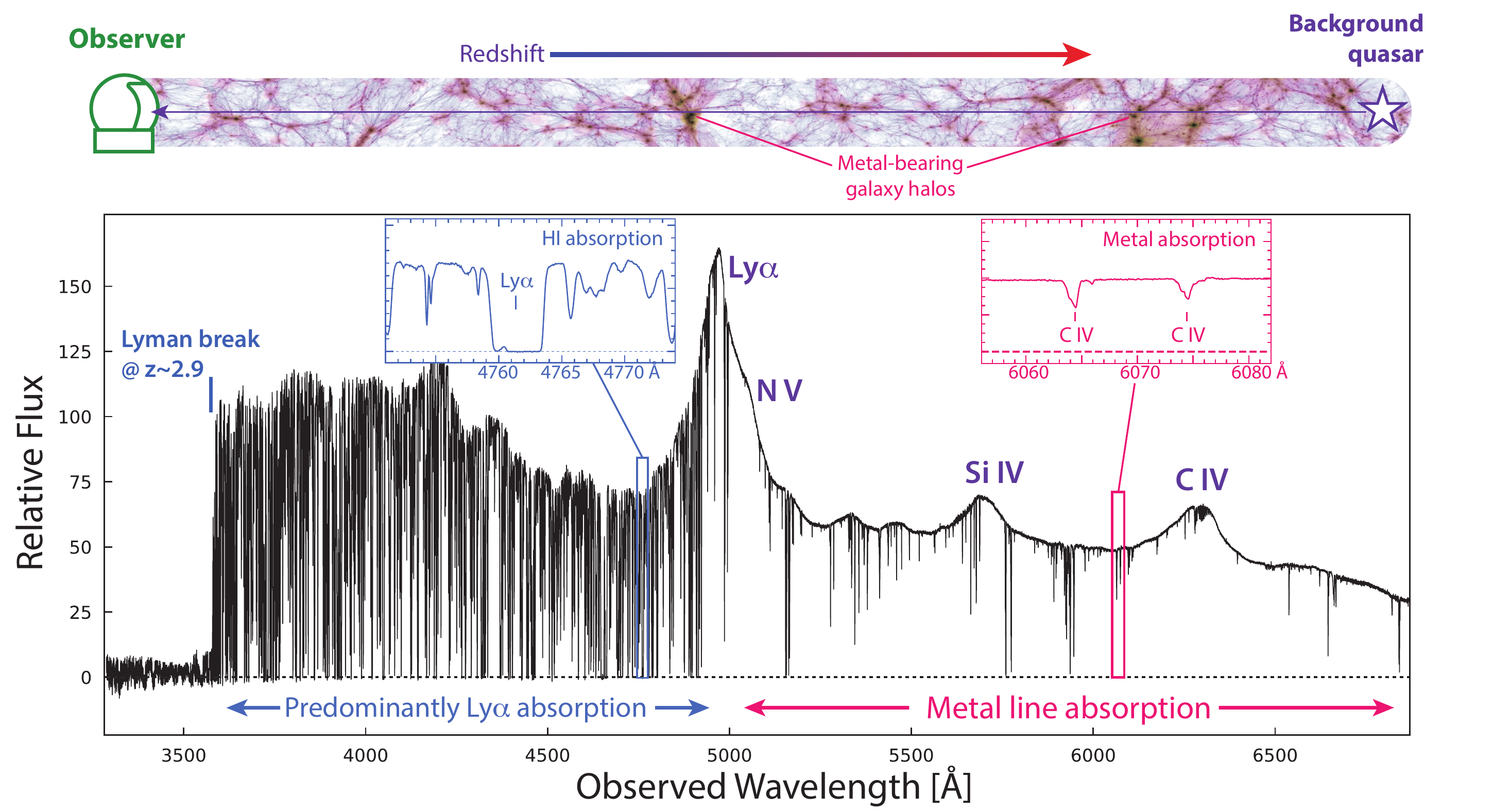}
\caption{An overview of using absorption line experiments to probe gas in the
Universe. {\it Top:} Light from a background object (in this case a quasar on
the right) follows a ray through the Universe toward an observer (left). As the
ray passes through the halos of galaxies (the darkest regions in the background
image), absorption due to \HI\ and metals is imprinted on the spectrum;
additionally, baryonic matter collected into the filaments will also absorb some
light, largely in the \HI\ Lyman-series transitions, giving rise to a forest of
absorption.
{\it Bottom:}
A sample high-resolution spectrum of a high-redshift ($z_{\rm em}$ = 3.0932)
quasar recorded at high resolution by UVES on the VLT
\citep{dodorico2016}. This spectrum has a ultra-high signal-to-noise ratio,
$SNR \approx 120$ to 500, with a spectral resolution $R \equiv  \lambda/\Delta \lambda
\approx 45,000$. Intrinsic emission lines from the quasar are marked across the
top. Many redshifted absorption lines are seen in these data. Those at
wavelengths longward of the quasar \lya\ emission all arise from metals; those
shortward of the \lya\ emission are largely from \HI, though with some
interloping metal lines. The absorption line profiles encode critical
information about the physical state of the gas. The insets show absorption
profiles from \HI\ (left) and representative metal lines (right) associated with
the $z\sim 2.9$ absorber producing the strong spectral break seen near 3600\AA .}
\label{fig:quasar_spectrum}
\end{figure}

At lower column densities, we designate systems with $1.6 \times 10^{17} \le
\NHI < 10^{19}$ \column\ as {\it ionized gas}. These are the Lyman-limit systems
\citep[LLSs;][]{tytler1982}, where the lower bound corresponds to an optical
depth at the Lyman limit $\tau_{912} \ge 1$. Observations and modeling of metal
lines from these absorbers demonstrate they trace ionized structures, with a
neutral fraction often ${\rm H^0/H} \approx 10^{-3}$ \citep[Figure
\ref{fig:sketch}e; cf.][]{fumagalli2016,wotta2019}.  These absorbers are easily
identifiable by their distinctive absorption break in quasar spectra (see
Figure~\ref{fig:quasar_spectrum}). These lower-density absorbers are found in
more extended regions around galaxies (Figure~\ref{fig:sketch}a, d) and are
subsequently observed with a higher frequency than the higher column density
systems (Figure~\ref{fig:sketch}b).

At even lower column (surface) densities, the gas becomes optically thin to
ionizing radiation and traces low-density circumgalactic and intergalactic gas.
The IGM manifests itself observationally
as a ``forest'' of \HI\ absorption lines in the spectra of background quasars,
with $10^{12} \le \NHI \le 1.6\times 10^{17}$ \column, the so-called {\it \laf}.
Its absorption is caused not by individual, confined clouds, but by a gradually
varying density field characterized by overdense sheets and filaments and
extensive, underdense voids that evolve with time \citep{mcquinn2016}.

\subsection{Basic Concepts of Absorption}
\label{sec:basic_abs}

As light from a distant source propagates to an observer, a fraction can be absorbed via electronic resonance transitions in intervening atoms, ions, or molecules and subsequently reemitted along another path. The specific flux, $F(\nu)$, observed from a background object is expressed in terms of the unabsorbed continuum flux,
$F_0(\nu)$:
\begin{equation}\label{eqn:i_eqn}
    F(\nu) = F_0(\nu)e^{-\tau(\nu)}.
\end{equation}
At rest frequencies $\nu$ corresponding resonance transitions in the intervening gas, the  optical depth, $\tau(\nu)$, due to absorption is related to the column density, $N$, of the absorbing species through
\begin{equation}
    \label{eqn:tau}
    \tau(\nu) = \int_0^{+\infty} n \sigma(\nu) \, ds = N \sigma(\nu),
\end{equation}
where $n$ is the density of absorbing particles and $s$ is the path through the material. The absorption cross section, $\sigma(\nu)$, is:
\begin{equation}
    \sigma(\nu) = \frac{\pi e^2}{m_e c} f \phi(\nu).
\end{equation}
Here $f$ is the oscillator strength describing the intrinsic strength of the
absorption line \citep[a recent tabulation of these ``$f$-values'' can be found
in][]{cashman2017}. The line profile, $\phi(\nu)$, is the intrinsic shape of the
aborption line with frequency, itself a convolution of two parts: a central
Gaussian core characteristic of the internal motions of the gas and broader
Lorentzian wings due to the intrinsic width of atomic energy levels (see Eqn.
6.37 of \citealt{draine2011}). For metal line absorption, the central Gaussian
component is the most important and is distributed such that $\tau(v) \propto
\exp(- v^2 / b^2)$ for velocity $v$ relative to line center and Doppler
parameter $b$. The Lorentzian wings become important for strong absorption,
typically only applicable to \HI. An observed spectrum, e.g., Figure
\ref{fig:quasar_spectrum}, is a summation of all of the optical depth profiles
-- some of which overlap -- as convolved by the instrumental line spread
function.

There are three principal approaches to relating the observed line profile to
column density (ultimately through Equation \ref{eqn:tau}). The first relies on a measurement of the restframe {\it equivalent width} of the absorption line, $W_{\rm rest}$, defined as:
\begin{equation}
W_{\rm rest} = \int \frac{F_0(\lambda) - F(\lambda)}{F_0(\lambda)} \, d\lambda_{rest}
    = \int (1 - e^{-\tau(\lambda)}) \, d\lambda_{rest},
\label{eqn:EW}
\end{equation}
where the restframe wavelength is related to the observed wavelength by the
absorber redshift, $\lambda_{\rm rest} = \lambda_{\rm obs}/(1+z_{\rm abs})$. The
column density of the absorbing species is then derived from measured equivalent
widths by the classical {\it curve-of-growth} technique
\citep[e.g.,][]{draine2011}.  The observed equivalent width is not dependent on
instrumental resolution, and this technique is useful when several
under-resolved lines are available  (e.g., using low-resolution spectroscopy).

An alternate approach integrates the {\it apparent optical depth} of an absorption line to derive its {\it apparent column density} \citep[][]{savage1991}. The apparent optical depth profile, $\tau_a(v)$, is derived directly from the normalized observed spectrum, inverting Equation \ref{eqn:i_eqn}. The ``apparent'' here allows for the fact that the observed flux has been convolved by the instrumental line spread function. The apparent column density, per unit velocity, is then
\begin{equation}
    N_a (v) = \frac{m_e c}{\pi e^2} \frac{\tau_a (v)}{f \lambda_0}
    = [3.768 \times 10^{14}\ {\rm cm^{-2}\, (km \ s^{-1})^{-1}}] \frac{\tau_a (v)}{f \lambda_0({\mathrm \AA})},
\label{eqn:Nav}
\end{equation}
where $\lambda_0 ( {\mathrm \AA} )$ is the central wavelength of the transition (in \AA). The apparent column density is an integration over this $N_a(v)$ profile.


The last approach to deriving column densities is {\it Voigt profile fitting,}
which models the absorption profile based upon the summation of the optical
depths from individual absorbing {\it components.}  A comparison of the instrumental line spread function convolved model
flux distribution with the observed spectrum constrains three parameters for each component: $N$,
$v$ (the central velocity), and $b$ (the Doppler parameter describing the
breadth of its Gaussian core). Ideally, the continuum is also included as a free
parameter to the fit. The most important concern of profile fitting is the non-uniqueness of the adopted component model. However, it provides component-by-component physical properties and accounts naturally for saturation when simultaneously using information from multiple transitions of an absorbing species.


Strong neutral hydrogen absorption with $\NHI > 10^{19}$ \column\ will show {\it
damping wings} in \lya\ due to the Lorentzian component of the optical depth
profile. The \HI\ column density is derived by fitting these wings (akin to the
profile fitting approach). Such strong absorption is on the {\it square-root
part} of the curve-of-growth, where the equivalent width grows as the square
root of the column. The column density of \HI\ can also be derived directly from
the equivalent width of \lya:
\begin{equation}
    \NHI =  (1.88 \times 10^{18} \ {\rm cm}^{-2} )
                \times W_{\rm rest}^2,
\end{equation}
where $W_{\rm rest}$ is given in \AA. Note that \cite{lee2019} discuss
departures of the fully quantum mechanical profile from the semi-classical Voigt
profile (approximated for a two-level atom) at very high column densities
(\NHI$>10^{22}$ cm$^{-2}$). Beyond spectroscopy, the cumulative absorption from
gas modifies the colors of background sources or provides a frequency-dependent
modification of light, providing a means to derive statistical properties
of the intervening matter \citep{prochaska2019, deharveng2019}.

\section{COSMIC EVOLUTION OF BARYONS}
\label{sec:baryon}

Multi-wavelength studies have accurately constrained the growth of galaxies
through cosmic time: the rate of star formation per comoving volume
peaks at a redshift z$\sim$2 (the ``epoch of galaxy assembly''), stays high up
to z$\sim$1 (Universe age: 2-6 Gyr), then dramatically decreases by more than an
order of magnitude from z$\sim$1 until today \citep{madau2014}. Probing the fuel
for star formation, i.e. the denser phase of the interstellar medium in
high-redshift galaxies, is essential to provide insights into the physical
processes producing these changes.

\subsection{Cosmic Evolution of Neutral Gas}
\label{sec:neutral}

Neutral atomic hydrogen (\hi) gas, a primary
ingredient for star formation, is therefore a key input to understand how
various processes govern galaxy formation and evolution. In this section, we
provide a review of recent measurements of the neutral phase of the gas in the
Universe and its evolution with cosmic time. We note that this census is more than an
accounting exercise, but rather aims at representing the cosmic cycling of
baryons \citep{fukugita2004, bouche2005, bouche2006, bouche2007, shull2012}.

\subsubsection{Basic Principles of Characterizing the Neutral Gas Mass Density}
\label{sec:basic_omega_neutral}

The column density distribution function (Figure~\ref{fig:sketch}b), the rate of
absorber incidence per unit column density per unit absorption path is
characterized as:
\begin{equation}
    f(N, z) dN dX = \frac{\mathcal{N}}{\Delta N \sum_{i=1}^{m} \Delta X_i} dN dX,
    \label{eqn_fN}
\end{equation}
where $\mathcal{N}$ is the number of absorbers observed in a column density bin $[N, N+\Delta N]$ obtained from the observation of $m$ background sources with total absorption distance coverage $\sum_{i=1}^{m} \Delta X_i$.

The distance interval, $\Delta X$, relates the observed redshift path to co-moving path length; thus, it depends on the geometry of the Universe. In the current preferred flat cosmology ($\Lambda$CDM), $X(z)$ is  \citep{bahcall1969}:
\begin{equation}
X(z)= \int_{0}^{z} (1 + z)^2 \left[\frac{H_0}{H(z)}\right] dz = \int_{0}^{z} (1 + z)^2 \left[\Omega_\Lambda + \Omega_m (1+z)^3\right]^{-1/2} dz,
\label{eqn_Xz}
\end{equation}
(assuming the curvature term $\Omega_k = 0$, and $\Omega_\Lambda$ is constant
with redshift). The comoving baryonic mass density of the gas is computed by
integrating the observed column density distribution:
\begin{equation}
\rhogas(z) = \frac{H_o \mu m_{\rm H}}{c }
                    \int_{N_{min}}^{N_{max}} N f(N,z) dN,
\label{eqn:rhogas}
\end{equation}
where $\mu$ is the mean molecular weight of the gas which is taken to be 1.3
(76\% hydrogen and 24\% helium by mass), $m_{\rm H}$ is the hydrogen mass, and
$N_{min}$ and $N_{max}$ are the bounds of the column density regime being
studied. The mass density is commonly expressed relative to the critical mass
density at z=0, \rhocrit, so that a non-evolving quantity results in a constant
value of the density, $\Omega(z)$:
\begin{equation}
    \Omega(z) = \rho(z) / \rhocrit = \rho(z)/(3 \, H_0^2/8\pi G).
\end{equation}
Here $G$ is the gravitational constant and $H_0$ the Hubble constant. For our
adopted cosmology, $\rhocrit = 1.36\times10^{11}\ M_\odot \ {\rm cMpc^{-3}}$.
Note that the comoving mass density we use here is related to the proper mass
density by $\rho(z) = \rho_{\rm proper}(z)/(1+z)^3$.

In practice the integral in Equation~\ref{eqn:rhogas} is sometimes fitted with
multiple power laws for the purpose of integration. These fits may diverge at
large column densities, forcing an artificial choice of cut-off ($N_{max}$ or
infinity). The choice can make a substantial difference in the resulting
\OmegaGas. An alternative is to utilize a 3-parameter Schechter function which
converges at high column densities \citep{peroux2003a, zafar2013b}. The
integration of the column density distribution indicates that the lowest column
density absorbers dominate by number, while the rare strong systems are major
contributors to the total neutral gas density \citep[e.g.,][]{peroux2003a,
noterdaeme2012}. It is a common misconception that such surveys are biased
against high column density material due to its relatively small cross section
on the sky. Absorption line surveys are ultimately probing the {\it volume
density} by assessing the column density per unit absorption path; they are
independent of the spatial distribution of the gas through characterization of
the absorber frequency in Equation \ref{eqn_fN} (see Figure \ref{fig:sketch}b).

Several distinct observational techniques have been used to measure the
abundance of neutral hydrogen in the Universe. The approaches change with
redshift due to the differences in transitions that are accessible. At
$z\la0.4$, the current generation of radio telescopes measures \HI\ 21-cm
emission, which provides a direct estimate of the \hi\ mass. The emission arises
from the hyperfine splitting of the ground state of neutral hydrogen into two
levels because of the spin-spin interaction between the electron and proton. An
electron located in the upper energy level can decay into the lower energy state
by emitting a photon with a rest wavelength of $\approx21$ cm or frequency
$\approx1420$ MHz.  Recent surveys have made robust measures of the local \hi\
mass function and mass density of galaxies \citep{jones2018,westmeier2018}. An
integral over these provides a measure of \OmegaGas. Observing at higher
redshift is challenging as the line is intrinsically faint and because of the
increased importance of radio frequency interference at low frequencies. An
important goal of the next decade will be to push the 21-cm observations
(through stacking or otherwise) to higher redshifts that overlap other
observational techniques, notably with SKA pathfinders such as ASKAP and MeerKAT
\citep{curran2018}.

Redshifts above $z\sim0.4$ are currently beyond the sensitivity of 21-cm
searches. Characterization of the \HI\ content at higher redshifts  rely on the
direct detection of  \HI\ \lya\ 1215 \AA\ absorption in the spectra of
background quasars. These observations yield direct estimates of \NHI\ for
neutral gas absorbers through the \lya\ damping wings. At $z\la1.7$ the \lya\
line is obscured by the Earth's atmosphere, requiring expensive space-borne
ultraviolet (UV) spectroscopy. A combination of the geometry of the Universe and
the paucity of absorbers requires sampling many sightlines. Limited blind
surveys have been done with HST and other single-object spectroscopic missions.
The most statistically robust surveys have made
use of preselection of candidate absorbers using strong metal-lines (typically
including \mgii) initially observed in ground-based surveys \citep[see][]{rao2017}.

Over the $1.7 \la z \la 5$ redshift range, the \lya\ transition is observed from
the ground; extremely large surveys of thousands of quasar absorbers have
brought such studies in a new era \citep[e.g.,][]{noterdaeme2012, bird2017,
parks2018}. These ambitious endeavors with 2.5-m class telescopes -- the Sloan Digital Sky Survey, SDSS,
in the northern hemisphere and the 2dF quasar survey in the southern
hemisphere -- advanced the field significantly primarily because they produced homogeneous advanced data products for well over one million low-resolution quasar spectra. In the near future, dedicated spectroscopic surveys on 4-m class telescopes will provide a
new wealth of low and medium-resolution quasar spectra in extremely large numbers, notably the WEAVE-QSO survey, the DESI experiment, and surveys with the 4MOST experiment. Clearly, such surveys
will require specific approaches to analyse these large data outputs.

At $z\ga5$, the strong \hi\ absorbers are challenging to identify
observationally, as the \laf\ becomes highly absorbed. The forest becomes opaque
for \hi\ neutral fractions $\ge 10^{-3}$, which provides a blanket of absorption
that leaves little continuum against which to search for strong \lya\
absorption.  There have been however recent success in measuring the \oi\ forest
\citep{becker2019} as well as the red damping wing's effect on the quasar flux
\citep{ davies2019}. Higher-resolution quasar spectroscopy with the next
generation of extremely large telescopes (e.g. ELT/HIRES, GMT/G-CLEF) will allow us to recover transmission peaks against which neutral absorbers may be measured up to the reionisation
redshift \citep{maiolino2013}.  Alternatively, observers will have to rely on
other tracers of neutral gas at the highest redshifts.

A promising approach for the future is intensity mapping, which uses low
spatial-resolution surveys of unresolved sources over large cosmological volumes
to map overdensities on large scales. \cite{kovetz2017} summarize current and
future intensity mapping experiments. A range of lines will be used to
statistically assess the densities of neutral gas traced by 21-cm emission
\citep{chang2010, masui2013} and molecular gas traced by CO \citep{keating2015}
and [\ion{C}{ii}] \citep{pullen2018} up to the highest redshifts, both from
existing observatories and from purpose-built future facilities: e.g.  CONCERTO
on the APEX telescope and AtLAST.

\subsubsection{Neutral Gas Mass Density}
\label{sec:omega_neutral}

\begin{figure}[t]
\includegraphics[width=5.in]{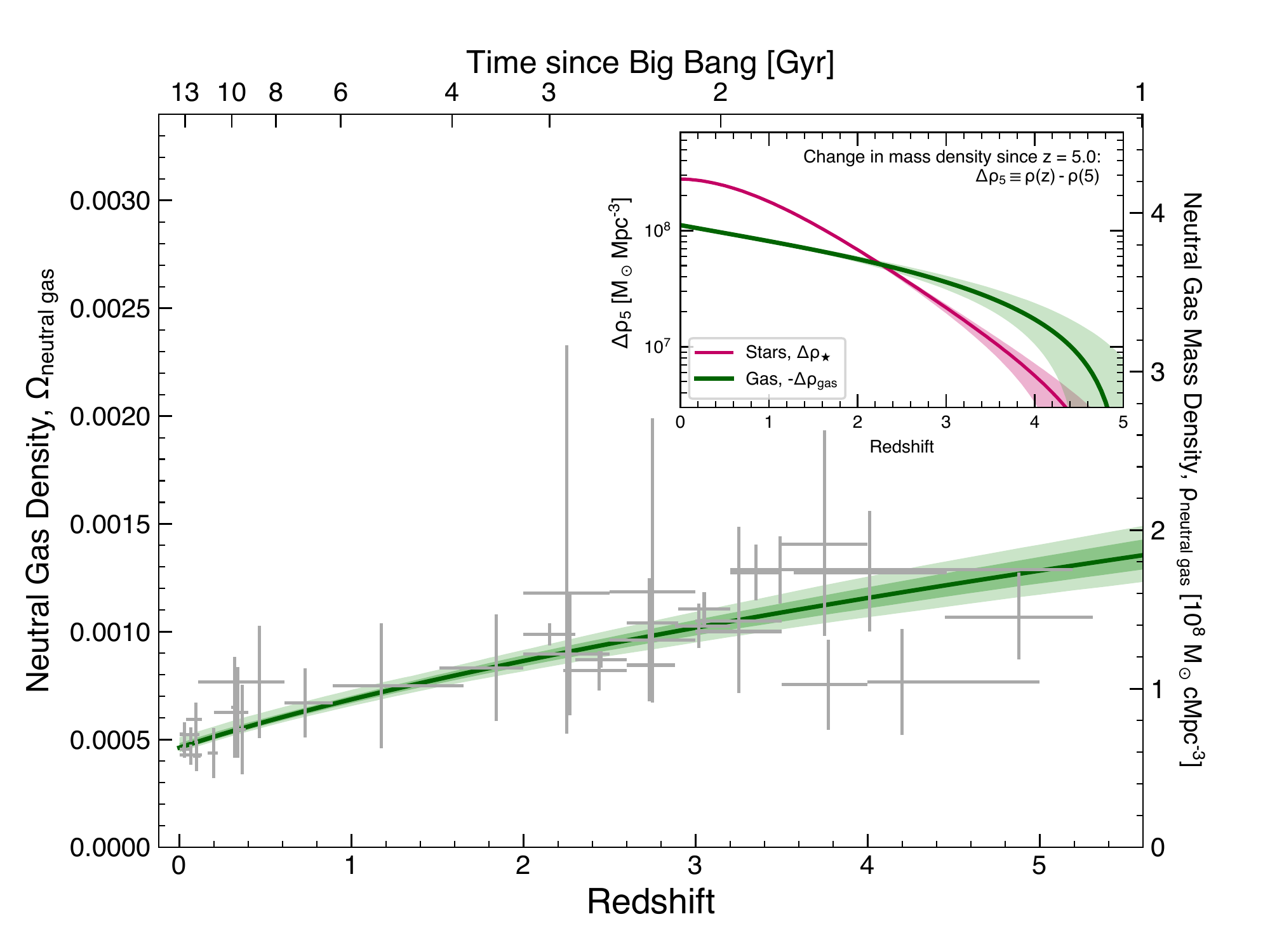}
\caption{Cosmic evolution of the neutral gas density, \OmegaGas$ \equiv
\rhogas / \rhocrit$.  In this and following figures, the corresponding comoving
density in mass units,  \rhogas, is shown on the right axis. The crosses display
the results from individual surveys from the literature. Those at $z\la0.4$ are
from 21-cm emission measurements; the higher redshift values are assessed
through \lya\ absorption.  The power-law fit described in Equation
\ref{eqn:powerlaw} is shown with the thick line. The gas density is on the
whole well determined and shows a mild evolution with cosmic time. {\it Inset:} The change in comoving gas mass density since $z=5.0$, $\Delta
\rho_{5}$.  By $z\la 2.5$, the total increase in stellar mass density exceeds
the neutral gas consumption, requiring replenishment of the neutral gas supply.  }
\label{fig:omega_neutral}
\end{figure}

Figure~\ref{fig:omega_neutral} shows a collection of \OmegaGas\ measurements
from the literature as crosses. We base this  compendium on results from the
last decade. Rather than try to be exhaustive, we chose representative results
that avoid redundant use of the same datasets.  Our compilation at low-redshift
($z\la0.4$) includes measurements from 21-cm emission surveys and spectral
stacking efforts \citep{delhaize2013, rhee2013, hoppmann2015, rhee2016,
jones2018, rhee2018, hu2019, bera2019}, although the latter is more prone to
confusion issues \citep[see][]{elson2019}.  Our  collated results at high
redshift include measurements using both high-resolution spectroscopic surveys
of tens of objects \citep{zafar2013b, crighton2015a, sanchez-ramirez2016,
rao2017} as well as analyses of lower-resolution spectra of thousands of objects
recorded by SDSS \citep{noterdaeme2009,noterdaeme2012}. We have homogenized the
data to our chosen cosmology and accounted for the contribution from helium,
multiplying the \HI\ values by 1/0.76. We note that 21-cm emission surveys
encompass the total \HI-mass, while absorption techniques reported here are
limited to systems with \lognhi$>$20.3. Lower column density quasar absorbers
contribute an additional 10--20\% to the total \HI\ content \citep{zafar2013b,
berg2019}.

\begin{marginnote}
\entry{Supplementary Material}{All of the data used to make the figures presented in this review are available as online supplementary tables. }
\end{marginnote}

In the absence of a physical motivation which would favour a specific functional
form, we fit the cosmological evolution of \OmegaGas\ with a mathematically
simple two-parameter power-law. We derive the best-fit function in $(1+z)$
space:

\begin{equation}
    \OmegaGas(z) = [(4.6\pm0.2)\times 10^{-4}] (1+z)^{0.57\pm0.04}
    \label{eqn:powerlaw}
\end{equation}
The 68\% and 95\% confidence intervals based on bootstrap estimates are shown in
Figure~\ref{fig:omega_neutral}. This fit is only appropriate at redshifts for
which we currently have data, since at the highest redshifts approaching
reionization the neutral gas density approaches the total baryonic density.

The comoving mass density of neutral gas evolves relatively little over time,
decreasing as $\propto (1+z)^{0.57}$ from $z\approx 5$ to today. Historically,
it was believed that the neutral gas provided all the raw material for the
formation of stars, so its redshift evolution should counterbalance the star
formation activity over the history of the Universe \citep{wolfe1986}.  The
relatively modest evolution throughout cosmic time of the neutral gas density is
in contrast to the now well-determined star formation rate density, which rose
sharply to a peak at redshifts z$\sim$1--2 before declining by an order of
magnitude to today's modest levels of star formation \citep{madau2014}. The
inset of Figure~\ref{fig:omega_neutral} shows the change in comoving mass
density since $z=5$, $\Delta \rho_{5}$, following \citet{crighton2015a}.  The
change in comoving stellar mass density is from an integration of the star
formation rate density of \citet{madau2014} (see \S \ref{sec:expected_metals};
we assume 50\% uncertainties). The negative change in neutral gas density is
based on the power-law fit in the main figure (with 68\% confidence interval
shaded).  At $z < 2.5$, the increase in stellar mass is significantly larger than
the neutral gas consumption. These observations call for the global and
continuous replenishment of the neutral gas from the ionized gas reservoir in
the intergalactic medium \citep[e.g.,][]{wolfe2005, prochaska2009,zafar2013b}.

From the theoretical viewpoint, the key processes of the production and
consumption of the neutral gas have to be introduced into simulations in the
form of sub-resolution effective models.  For these reasons, simulations of the
cold phase of the gas have relied on so-called semi-analytical approaches
\citep{lagos2011, dave2017}.  More recently, IllustrisTNG simulations
\citep{nelson2019} in particular have been post-processed to reproduce the
cosmological evolution of neutral gas \citep{diemer2019}.  Improved methodology
to model the atomic-to-molecular transition on the galaxy-by-galaxy basis
(including empirical, simulation-based, and theoretical prescriptions) have been
used to compare the cold gas mass function with z=0 observations. Reliable predictions will have to await more complete treatment of
self-shielding and the full chemistry in higher-resolution cosmological
simulations, which to date remain computationally challenging
\citep{richings2014b, maio2015, emerick2019}.

\subsection{Cosmic Evolution of Molecular Gas}
\label{sec:omega_mol}

The molecular phase is at the heart of the physical processes through which gas
is converted into stars. Neutral hydrogen provides the essential fuel, but this
fuel has to cool and transform to the molecular phase for star formation. It is
now possible to constrain the molecular content of large sample of galaxies
\citep{saintonge2011b, tacconi2020}.  The first unbiased survey for molecular
gas done through a blind scan was by \cite{walter2014}.  An analogous blind
survey, ASPECS, has been done in ALMA 3 and 1 mm (respectively $\sim$100 and 250
GHz) bands in the Hubble Ultra Deep Field \citep[UDF,][]{decarli2016,
decarli2019}.  ASPECS is the first assessment of the cosmic evolution of
molecular gas to $z<4$; the results are subject to significant uncertainties due
to small number statistics and the impact of cosmic variance.   In addition, the
VLA-based COLDz survey made complementary measurements at z$>$4
\citep{riechers2019}, while the IRAM 30m-based xCOLD GASS provided robust z=0
measurements \citep{saintonge2017}.  Figure \ref{fig:omega_all} illustrates our
current knowledge of the cosmic evolution of the molecular gas from these
surveys. We note that CO absorption provides an alternate approach to assessing
\OmegaMol; \citet{klitsch2020} use the ALMA
calibrator archive to search for absorption, providing stringent limits on the
molecular gas density at $z<1.5$ based on non-detections of CO. These approach is promising has it allows to reach lower gas column density free from cosmic variance issues \citep[see also ][]{kanekar2014}.

The molecular measurements reported here arise from studies undertaken in just
the last few years, we expect that observational determination of \OmegaMol\
will make rapid progress in the decades to come. In addition to improve sample
size and observation depth, there are some limiting systematic issues in using
CO as a tracer of molecular gas that will require a better theoretical
understanding to improve. The limited simultaneous frequency coverage
(bandwidth) of current receiver arrays produce a redshift-transition degeneracy
inherent to single-line detections.  Moreover, the observed higher rotational
quantum number transitions require a conversion to the reference CO(1--0). This
step introduces an uncertainty due to variation in poorly known Spectral Line
Energy Distribution (SLED) of different galaxy types \citep{carilli2013,
klitsch2019}.  The  CO-to-H$_2$ conversion factor adds a large systematic
uncertainty, as well.  Indeed, CO likely breaks down as a reliable tracer for
H$_2$ mass in extreme environments.  In particular, the conversion factor
transforming CO intensity to H$_2$ mass increases with decreasing metallicities
as CO photo-dissociates to a larger depth in each cloud, and as a result varies
both from object to object as well as with redshift \citep[e.g.][]{bolatto2013}.
Other tracers may eventually provide feasible substitutes for CO searches, such as [\ion{C}{i}] and [\ion{C}{ii}] \citep{papastergis2012}. In
principle, the direct measurement of resonant electronic Lyman and Werner bands
of H$_2$ in absorption  at UV-wavelengths in \hi\ absorbers \citep[found in $\sim4\%$ of absorbers;][]{bolmer2019} could be used to explore the scalings between CO, \ion{C}{i}, and \Hmol\  \citep{noterdaeme2018}.

On the theoretical side, the molecular phase of the cold gas is challenging to model because of the complex unknown physics involved. In addition, modelling below the hydrodynamic simulation's resolution (coined ``sub-grid'' modelling) is required to capture this phase of the gas.
Simple semi-analytical techniques based on metallicity and pressure-based models
\citep[][and references therein]{krumholz2013} are used to split the cold hydrogen from hydrodynamics
simulations (such as EAGLE or Illustris) into its atomic and molecular
components \citep{lagos2015,popping2019}. Simulating the neutral and molecular phases of the cold gas in full cosmological context is therefore one of the most crucial and challenging objective of studies of galaxy formation and evolution in the decades to come.

\subsection{Cosmic Evolution of Condensed Matter}
\label{sec:star_cold_gas}

Having summarized the densities of cold gas, including the neutral (\S\ref{sec:omega_neutral})  and molecular (\S\ref{sec:omega_mol}) components, we combine these with measurements of the stellar mass density to study the transitions between these baryonic components of the Universe. We refer to these components collectively as {\it condensed} matter.
Our goal is to assess the baryon inventory of condensed material and its cosmic
evolution over a look-back time of $\sim$ 13 Gyr. Building on previous works
that have performed such accounting exercises \citep[e.g.][]{pettini1999,
fukugita2004, bouche2007, shull2012}, we here present the change of the
different components with time as gas is converted into stars \citep{putman2017,
driver2018}.

Understanding the efficiency with which baryons are converted into stars is a
challenge in studies of galaxy formation and evolution. The evolution in the
star formation rate density is well established from observations of
star-forming galaxies across cosmic time at infrared, ultra violet,
submillimetre, and radio wavelengths \citep{madau2014}.  The star-formation rate
(SFR) density increased at early times, reached a peak at around $z\sim2$, and
subsequently decreased through today. Identifying the physical processes  that
drive this dramatic change and their relative importance  are active areas of
current research. 

\begin{figure}[h]
\includegraphics[width=5in]{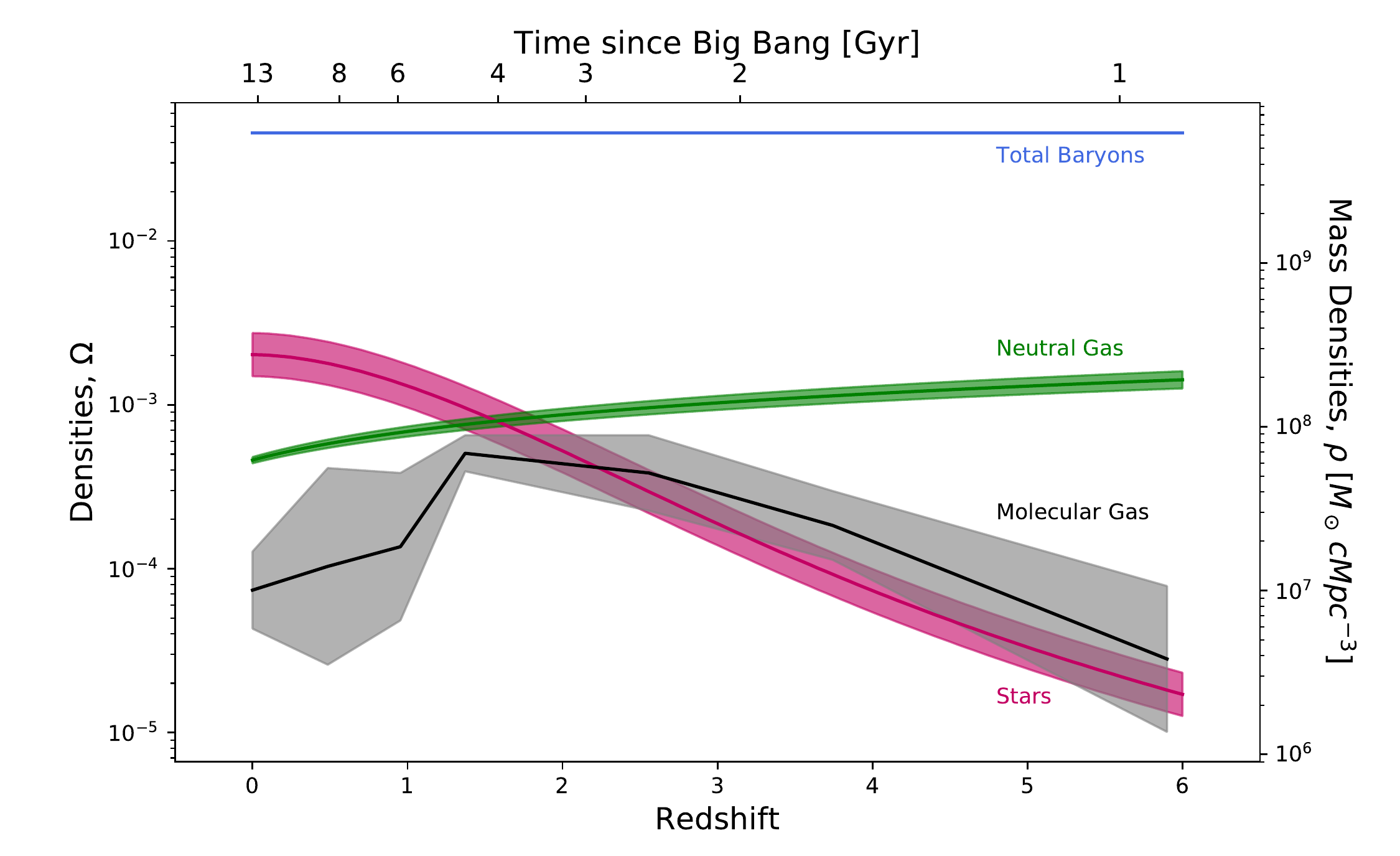}
\caption{Cosmic evolution of the density in condensed matter (stars, neutral
gas, molecular gas). The total baryonic density of the Universe estimated from
CMB anisotropies and from light element abundances is shown at the top.  The condensed
forms of the baryons illustrate the cycling of baryons most closely contributing
to the make-up of galaxies.  At all redshifts the neutral gas density dominates over that of the molecular gas. The molecular
gas density evolution roughly mirrors that of the star formation rate density
\citep{madau2014}, hinting that the molecular gas reservoir drives the evolution of star formation.  } \label{fig:omega_all}
\end{figure}

Figure~\ref{fig:omega_all} summarises the observable quantities making up the
baryon content of the ``condensed'' phases of the Universe, those collected into
high density regions. At the top of the plot, the horizontal line
indicates the total baryon density of the Universe. Both measurements of CMB anisotropies \citep{planck2016} and light element abundances coupled with Big Bang nucleosynthesis \citep{cooke2018} lead to $\OmegaBaryons \approx 0.0455$ (in our adopted cosmology).  The fact that these two independent measures from vastly different physical processes occurring at enormously different cosmic ages agree to within 2-$\sigma$ is a real triumph of cosmology. The mass in stars is well constrained from
measurements of the stellar mass density (\S\ref{sec:expected_metals}), and it
is found to build up steeply with redshift.  The neutral phase of the gas has
also been well constrained in recent years; we show the fit to the data from
Figure \ref{fig:omega_neutral}, indicating a steadily decreasing density with
cosmic time (\S\ref{sec:omega_neutral}).  We also plot the molecular gas density
from the blind CO emission line surveys discussed in \S\ref{sec:omega_mol}.

Globally, Figure~\ref{fig:omega_all} shows the neutral gas density dominates the
molecular gas density at all redshifts. This differs from the assumptions on galaxy scales of \citet{tacconi2018}, who argue that neutral gas in high redshift galaxies is negligible \citep[see][]{tacconi2020}. The molecular gas and stellar components have similar densities until $z\sim2$, after which the molecular gas density decreases rapidly, while the stellar mass of the Universe continues to increase.
Overall, the shape of the molecular mass density mirrors that of the star
formation rate density of the Universe \citep[][]{decarli2019, tacconi2020}. Our findings indicate that the history of star formations rates is primarily driven by the cosmic evolution of the gas reservoir.

In this review, we elected to focus on the densest baryons (the condensed matter
in our nomenclature); these are the most strongly coupled to star formation
within galaxies and are the most robustly observed. Naturally there are a number
of other contributors to the baryon budget which are either challenging to probe
observationally and are not included in Figure~\ref{fig:omega_all}.  At high-redshift, cosmological hydrodynamical simulations indicate most of these baryons are in the low-density ionized gas of
the \laf, for which deriving the density from observations is highly
model-dependent. At lower redshifts, a larger fraction of baryons are collected
on the galaxy halo scale, including in the CGM \citep{werk2013, tumlinson2017}.
A major fraction of the baryons at lower redshift is found in hot gas
(10$^5$-10$^6$K), including the gas in groups and clusters as well as the warm-hot intergalactic medium \citep[WHIM;][]{cen1999, dave2001}. This gas can be probed in X-ray experiments and more recently through the distortions the gas imprints on the CMB spectrum due to thermal and bulk
motions of free electrons \citep[the thermal and kinetic Sunyaev-Zel'dovich effects;][]{lim2018, de-graaff2019, tanimura2019}.

\begin{figure}[h]
\includegraphics[width=4in]{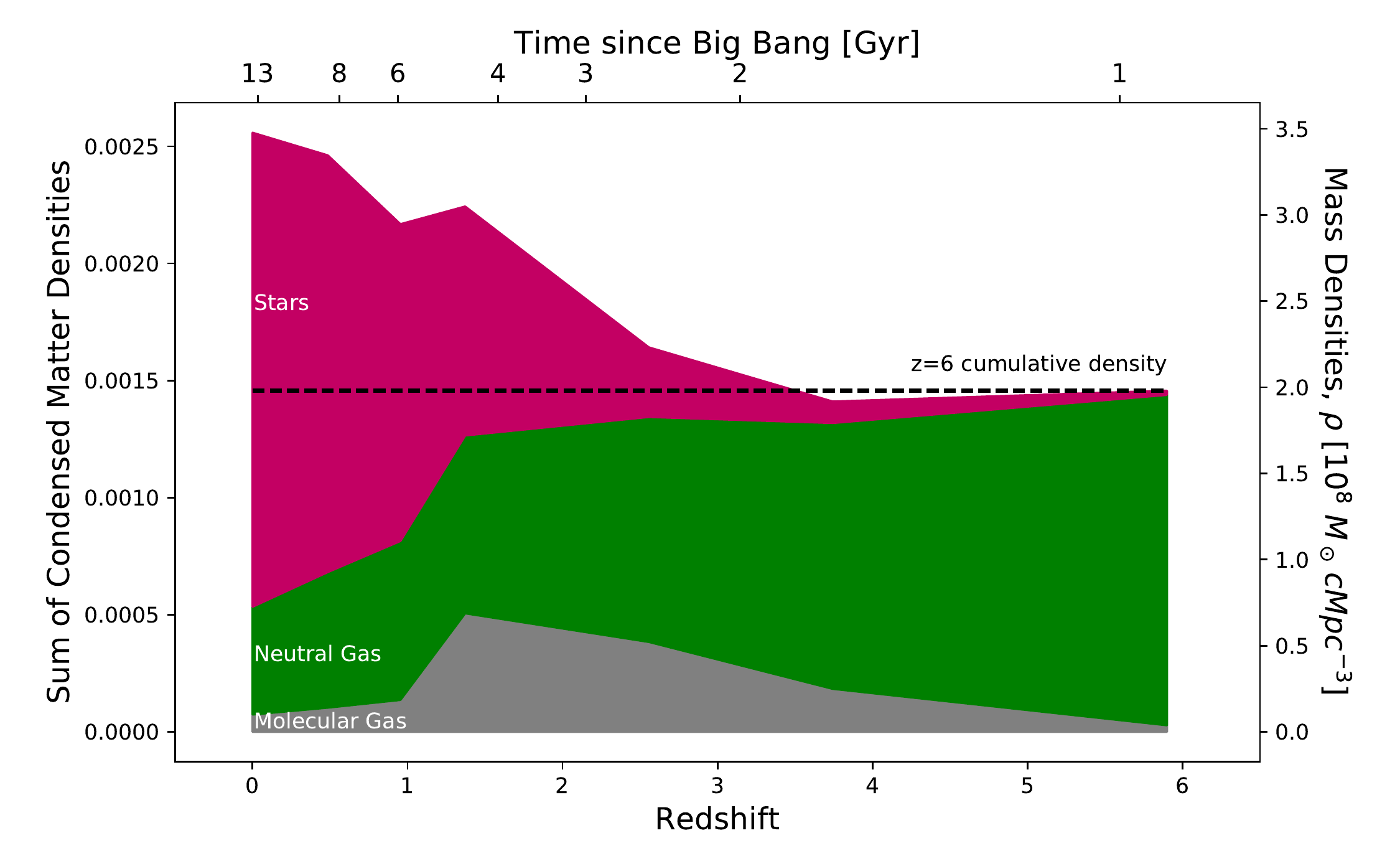}
\caption{Cumulative view of the redshift evolution of the total density of condensed matter: summing the molecular gas, neutral gas, and stars. At z$>$3, we witness the formation of \Hmol\ gas from \hi\ resulting in a constant amount of condensed matter (horizontal dashed line). At z$<$3, the amount of neutral gas decreases and the molecular gas is rapidly consumed into star formation. }
\label{fig:omega_cond}
\end{figure}

We can use the combined masses of the neutral (\S \ref{sec:omega_neutral}) and
molecular gas (\S \ref{sec:omega_mol}) to trace the conversion of gas into stars
over cosmic time.  In the broadest terms, \hi\ clouds form, and these molecular
clouds then cool, fragment, and initiate star formation in galaxies. Thus, the
cold gas (neutral+molecular) is expected to provide the reservoir of fuel
for star formation activity. Figure \ref{fig:omega_cond} provides a cumulative
view of the observed densities of these dense baryonic phases. Focusing on the
total, we note an increase in the combined density with time, indicating that
the cold gas reservoir at early times is insufficient to provide for all of the
stellar and cold gas content seen today (comparing the sum at lower redshift
with the dashed line denoting the value seen at the highest redshifts).  At
$z\ga2.5$, the total density of condensed matter is relatively constant; we see the
conversion of \hi\ into \Hmol\ and stellar mass. However, at lower redshifts the
\hi\ is rapidly depleted into molecular gas, while the stellar mass density
grows at an even higher pace. At $z\le2.5$, the combined neutral and molecular gas
densities decrease strongly, though the total mass in condensed matter grows,
reaching nearly twice its value at the highest redshifts. This requires a
continues conversion of other forms of matter into these condensed forms in
order to provide for the overall growth.

\begin{figure}[h]
\includegraphics[width=3.4in]{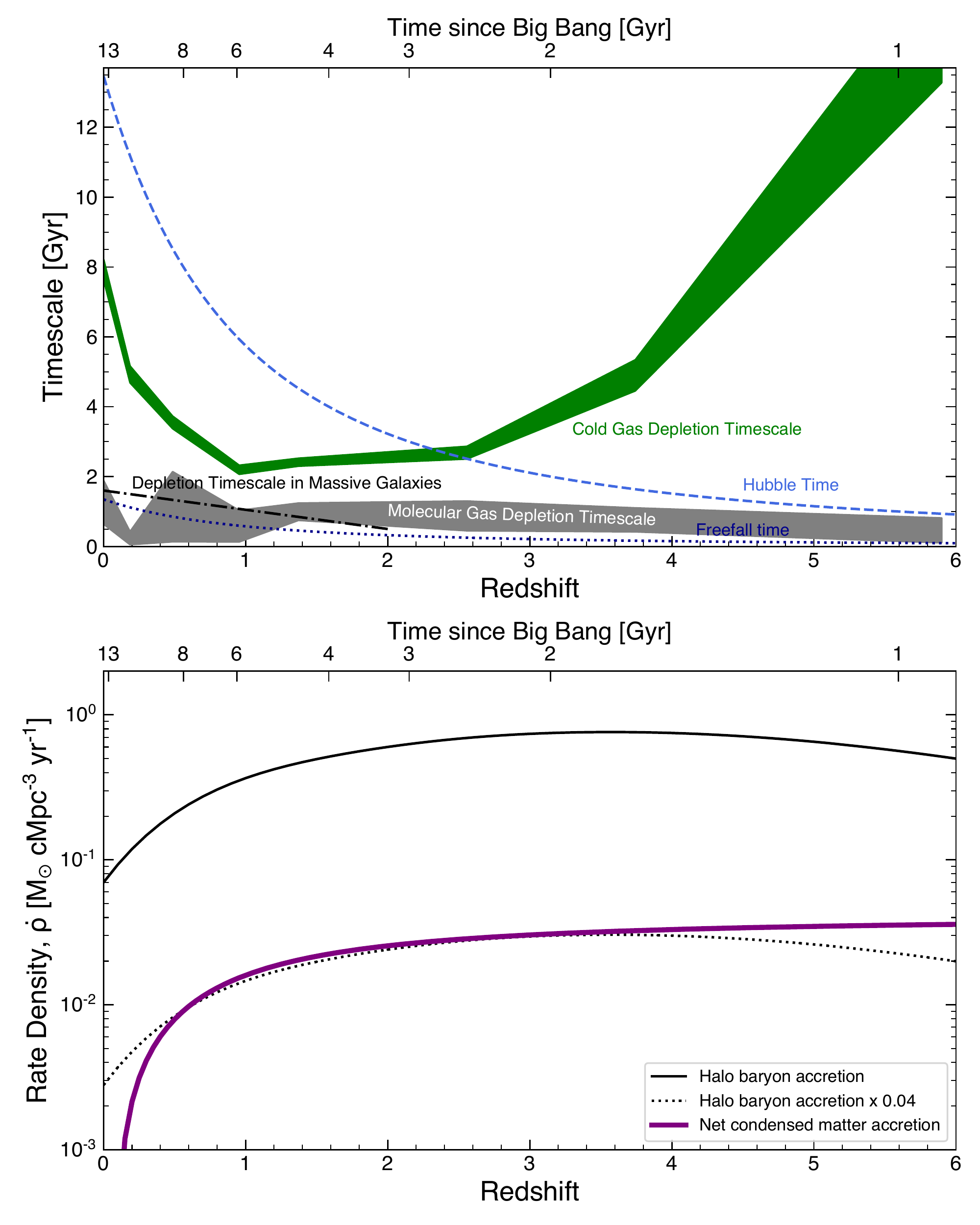}
\caption{Top: Timescales for conversion of gas in its various forms to stars at
each redshift. The filled areas show the depletion timescales of the
molecular or cold gas. The molecular gas depletion scale shows a
constant value with redshift, whereas the cold gas depletion time
(inclusive of molecular and neutral gas) varies significantly with
redshift. The dot-dashed line shows the functional fit to the
molecular depletion timescale measured in a sample of massive galaxies
by \cite{tacconi2020}, which is in agreement with our global
estimate. The cold gas depletion times are shorter than the dynamical
time (taken to be 10\%\ of Hubble time), implying the need for
conversion of ionized gas from the circumgalactic medium and/or
intergalactic medium to supply additional fuel for star formation. At
z $\la$ 1, the molecular gas consumption time is comparable with the
dynamical time, implying the reservoir of molecular gas is slowly
becoming insufficient to fuel the observed star formation. Bottom: The
purple line shows the net accretion rate density required to reproduce
the observed evolution of the mass density in condensed matter (Figure
\ref{fig:omega_cond}). Also plotted is the baryonic accretion rate
onto halos as a function of redshift. In addition, we show the
baryonic accretion rate density scaled by an efficiency 0.04 for
conversion of accreted baryons into condensed matter, showing the
similarity between the shape of the accretion rates that are
continuously decreasing with time, dropping significantly at z $<$
1.5.  Together, these results indicate that globally the growth of
condensed matter (stars and cold gas) in the Universe scales
principally with the dark matter accretion rate onto halos. The gas
regulator model, in which star formation in galaxies is
instantaneously regulated by the mass of the gas reservoir \citep[e.g.,][]{bouche2010,lilly2013}, is in remarkable agreement with
this global picture of a continual cycle of baryons flowing in and out
of galaxies as a key moderator of galaxy evolution. 
}
\label{fig:gas_depletion}
\end{figure}

A measure of the rate of gas conversion into stellar mass is the gas depletion timescale, the time that would be required to convert the gas into stars at the current rate. It is expressed as:
\begin{equation}
    \tau_{\rm dep}=\rho_{\rm gas}/\dot{\rho}_\star,
\end{equation}
where $\dot{\rho}_\star$ is the star formation rate density.
\Tdep\ is therefore related to the Kennicutt-Schmidt relation \citep{kennicutt1998}.
The inverse of \Tdep\ is often defined as the star formation efficiency,
SFE=1/\Tdep, even though it is related to a rate, expressed in units of yr$^{-1}$. Figure \ref{fig:gas_depletion}
presents a comparison of the molecular and cold (neutral+molecular) gas
depletion timescales as function of redshift.  The allowed values of the depletion time range from \Tdep$ = 100$ Myr to 1.5 Gyr, in line with previous findings \citep{saintonge2017, tacconi2018,
riechers2019}. \Tdep\ is roughly constant with redshift, within the sizeable uncertainties (denoted by the filled
area in Figure~\ref{fig:gas_depletion}).
\cite{tacconi2020} find a mild redshift
evolution in \Tdep\ measured in massive galaxies, proportional to (1+z)$^{-1}$ (dashed-dotted line in
Figure~\ref{fig:gas_depletion}). While coming from different approaches
(measurements of the global quantities of baryons in the Universe versus in-situ
observations targetting galaxies), the two results are remarkably
consistent. It confirms that globally the molecular gas depletion timescale depends weakly on stellar mass,
which is consistent with recent observations at low redshifts \citep[e.g.,][]{tacconi2018}. The steady timescale for global molecular gas depletion over
$\sim$13 Gyr lookback time points toward a fundamental and universal physical
process of conversion of molecular gas into stars. This is in contrast to
$\tau_{\rm dep}$ for cold gas (neutral+molecular), which is large at high redshifts. We compare these values to the typical dynamical time for halos in Figure~\ref{fig:gas_depletion}, which is roughly $\sim$10\%\ of the Hubble time \citep{lilly2013, tacconi2020}. We note that the cold (neutral+molecular) gas depletion timescale is much longer than the dynamical time at all redshifts, indicating that the reservoir of cold gas is sufficient to sustain the rate of star formation for many dynamical times at all epochs. At z $\la$ 1, the molecular gas consumption time is comparable to the dynamical time, implying the reservoir of molecular gas is slowly becoming insufficient to fuel the observed star formation. There is a large reservoir of cold gas at these redshifts, but the efficiency with which it is converted into molecular gas is low.

The increase in the cumulative mass densities of cold gas and stars with decreasing redshift calls for the intake of new material into condensed matter. From this we infer the {\it required} rate of transformation to reproduce the rise in total condensed matter. This is the result of the {\it net} conversion of ionized gas from the CGM and IGM reservoirs into cold gas and stars \citep[e.g.,][]{wolfe2005, prochaska2009, bauermeister2010, zafar2013b}. This is inclusive of recycling material on galactic scales, with some material previously ejected and ionized by feedback returning to the condensed phases. We plot this net rate of conversion of ionized gas into
the condensed phase per unit volume in Figure~\ref{fig:gas_depletion}. The net accretion rate density continuously
declines with cosmic time. Also plotted is the baryonic accretion rate onto halos as a function
of redshift. We calculate the total mass accretion as the virial mass growth rate of halos from
\cite{dekel2013} integrated over the \cite{bocquet2016} mass functions at each redshift. The
baryon mass accretion rate is the total matter accretion rate scaled by the cosmic baryon fraction ($\Omega_b$/$\Omega_m$ = 0.15 in our adopted cosmology). The baryon and net condensed matter accretion rates
have similar shapes; both are high until z $\sim$ 1.5 and then drop significantly through today. The
drop in accretion rate at low redshift is caused in part by the expansion of the Universe as galaxies
are driven away from each other by the repulsive force of dark energy; accretion and merging
is slowed, and galaxies are gradually starved of incoming fuel. This accretion rate includes the
effects of both smooth accretion and mergers, whose mix changes with redshift and halo mass \citep{padmanabhan2020}.
We also plot in Figure~\ref{fig:gas_depletion} the baryonic accretion rate density scaled by an efficiency of 0.04
for conversion of accreted baryons into condensed matter, showing the similarity between the
shape of the accretion rates. These similarities point to a causal relation between dark matter
accretion and the rate of growth of condensed matter. This efficiency describes the conversion of
accreted baryons into condensed matter for material accreting onto all halos—including cluster
scales. The low efficiency rate implies that a majority of the new baryons accreted into halos
are in ionized forms, both warm photoionized material and hot coronal gas. There are several
reasons for this. Gas being accreted onto halos from the IGM has a low-enough density that it
is necessarily ionized. Furthermore, the accretion process involves shocks that heat the incoming
gas significantly, especially as matter is accreted onto massive galaxies, groups, and clusters. At the
same time, we are comparing the baryon accretion rate with the net rate of growth of condensed
matter. The effects of feedback and galactic winds will be to convert some of the cold gas into
ionized material. Even given all of these baryonic processes, globally the growth of condensed
matter (stars and cold gas) in the Universe scales principally with the dark matter accretion rate
onto halos. These results indicate that the z $<$ 2 decline of the star-formation history is driven by
the lack of molecular gas supply due to a drop in net gas accretion rate, which is itself driven by a
decrease in the dark matter halos accretion rate.
At early times (z $\ga$ 2), the global SFR density of the Universe increases as the baryon net accretion
rate stays high. At late times, the gas reservoir depletes as the accretion rate significantly
decreases, and galaxies are quenched. These results are in line with expectations from the gas
regulator (or bathtub) model \citep[e.g.,][]{bouche2010, dave2012, lilly2013}, which
uses continuity equations to describe the cycling of baryons in and out of galaxies. The regulator
model describes galaxies as systems in a slowly evolving equilibrium between inflow, outflow, and
star formation. At early times, galaxies build up baryonic matter during an epoch of gas accumulation,
in which the SFR is limited by the available reservoir of cold gas. At later times, galaxies
reach a steady state in which global star formation is regulated by the net accretion rate. Given
its simplicity, this gas regulator model is in remarkable agreement with the global picture presented
in Figure~\ref{fig:gas_depletion}, in which the continual cycle of baryons flowing in and out of galaxies is a key
moderator of galaxy evolution. (See the sidebar titled Cosmic Evolution of Baryons.)

\begin{textbox}[h]
\section{Cosmic Evolution of Baryons}
\label{box:omega_cond}
A modern census of \OmegaGas\ indicates that the cosmic evolution of \hi\ is now observationally well constrained. Results indicate a mild evolution with a decrease proportional to (1+z)$^{0.57}$. It will be important that 21-cm emission measurements reach the epoch where \lya\ absorption measurements against background quasars are also available to allow for a direct comparison of these two vastly different observing methods. Intensity mapping provides a promising avenue to further push these measurements to the highest redshifts.

While still in their infancy, new measurements of \OmegaMol\ allow us to make a global assessment of the condensed matter (cold gas and stars) in the Universe from direct observables. At $z>2.5$, the formation of \Hmol\ from \hi\ results in a nearly constant amount of cold gas. At lower redshift, \Hmol\ decreases rapidly, following the trend in star formation rate. The \Hmol\ depletion timescale is constant with redshift, suggesting a universal physical process of conversion of molecular gas into stars on global scales. At z $\la$ 1, the molecular gas consumption time is comparable with the
dynamical time, implying the reservoir of molecular gas is slowly becoming insufficient to fuel the observed star
formation. At z $\la$ 3, the increase in the total density of condensed material requires accretion of the ionized phase
of the gas from the CGM and/or IGM baryon reservoir.

We find that the baryonic accretion rate density onto dark matter halos scaled by an efficiency factor of 0.04
shows similarity with the shape of the net condensed matter accretion rate, which are continuously decreasing with
time and dropping significantly at low redshift. These results indicate that the z $<$ 2 decline of the star-formation
history is driven by the lack of molecular gas supply due to a drop in net gas accretion rate, which is itself driven
by the decreased growth of the dark matter halos. These observations are in remarkable agreement with the gas
regulator model, which describes a continual cycle of baryons flowing in and out of galaxies as a key moderator of
galaxy evolution.

\end{textbox}

\section{COSMIC EVOLUTION OF METALS}
\label{sec:metal}

The detection of cosmic matter through absorption lines is also key to trace the metallicity evolution of the Universe. This is done by measuring absorption from metal lines arising in the same gas used to trace the hydrogen gas content of the Universe (e.g., \S \ref{sec:omega_neutral}). Using absorption against background sources allows us to assess the metallicity evolution of the Universe with similar sensitivity over $0 \la z \la 5$; it provides a measure of the metallicity weighted by the total gas mass of the Universe rather than by stellar luminosity, as do most metal tracers (stars and \hii\ regions). Absorption line-based metallicities are independent of excitation conditions -- they are largely insensitive to density or temperature and require no local source of excitation. This differs from emission line metallicity estimates, where density and temperature information is critical to making high-quality measurements \cite[see e.g.][]{maiolino2019, kewley2019}. The empirical calibrations used to circumvent the need for physical conditions for most studies yield disparate results \citep{kewley2008}. Absorption lines, on the other hand, probe low- and high-metallicity as well as low- and high-excitation regions independently of galaxy stellar masses. In this section, we review the modern estimates of metallicity in various environment and make a census of the total amount of metals in the Universe.

\subsection{Basic Principles of Measuring Metallicity through
Absorption Lines}
\label{sec:basic_metal}

The {\it metallicity} of astrophysical matter summarizes the abundance of metals (elements greater than helium) present relative to hydrogen by number; we express the logarithmic abundance relative to solar of a specific element X as
\begin{equation}
[{\rm X/H}] = \rm \log \{ N({\rm X})/N({\rm H}) \} -
                \log \{ N({\rm X})/N({\rm H}) \}_{\odot},
\end{equation}
where $N({\rm X})$ refers to the column density (surface density in atoms cm$^{-2}$) of that element, and $\{ N({\rm X})/N({\rm H}) \}_{\odot}$ is the reference abundance in the Solar System. Throughout this review we adopt the Solar System abundances summarized in \citet{asplund2009}.

\begin{marginnote}[]
    {\bf Logarithmic abundances relative to the Solar System:}
    \entry{[X/H]}{abundance of a specific element X}
    \entry{[M/H]}{abundance of all metals M}
\end{marginnote}

Thus the metallicity $[{\rm X/H}]$ is the logarithmic abundance relative to the assumed solar abundance. When probing regions dominated by neutral gas, the measurements  probe the dominant ionization states of X and H, which implies assuming
\begin{equation}
N({\rm X}) = N (\mbox{\ion{X}{ii}}) \ {\rm and} \ N({\rm H}) = \NHI.
\end{equation}
There are important exceptions, e.g., the dominant ionization phases are the
neutral states for N and O.  Similarly, one calculates relative abundances of
metals, e.g., [X/Y]. This comparison provides insight into the relative
nucleosynthetic contributions of the elements (e.g., \S
\ref{sec:ionization}) and depletion of the elements into the solid phase
(dust; \S \ref{sec:depletion}).

\subsection{Challenges in Assessing Metallicity}
\label{sec:challenge}

Absorption line techniques provide a means of tracing the cosmic metal evolution. The column densities of the gas is measured with high precision. For ionized and partially-ionized gas, the precision of the derived metallicities is limited by the need for ionization corrections (\S \ref{sec:ionization}). In all absorbers, it is also affected by dust depletion (\S \ref{sec:depletion}) and dust sample bias (\S \ref{sec:samplebias}). The details of these effects and corrections to account for them are discussed below.

 \subsubsection{Ionization Effects: Accounting for Differential Ionization}
 \label{sec:ionization}

The gas traced by absorption lines is often multi-phased, with sight lines passing through regions of differing temperature, density, and ionization conditions. Observations measure atomic hydrogen and a fraction of the metal ionic states. The presence of gas with neutral H fractions significantly less than unity requires ionization corrections to transform a measured ratio of ionic to \HI\ column density $N(X^i) / N(\Hatom)$ into $X / {\rm H}$. Methodologically, we write the abundance $X / {\rm H}$ as
\begin{equation}\label{eqn:icf}
    \frac{X}{{\rm H}} = \frac{N(X^i)}{N({\rm H}^0)} \frac{\xHI}{x(X^i)},
\end{equation}
where $x(X^i) \equiv N(X^i)/N(X_{\rm total})$ is the ionization fraction of
$X^i$ and the ratio of ionization fractions in the last term is the {\it
ionization correction factor}, $\ICF(X^i) \equiv {\xHI}/{x(X^i)}$. Even systems
with high ionization fractions can have $\ICF(X^i) \approx 1$ for specific
choices of $X^i$. The ionization corrections range from unimportant to more than
an order of magnitude in partially-ionized \citep{quiret2016} and ionized gas
\citep{fumagalli2016, wotta2019}. Ionization corrections are negligible in
strongly-neutral regions \citep[typically with $\logNHI \ga
20.3$; ][]{vladilo2001} with the exception of a few atomic or ionic species typically observed only in the densest gas (e.g., \ion{C}{i}, \ion{Na}{i}, \ion{Ca}{ii}).

The models used to estimate ICFs employ sophisticated atomic physics and
radiative transfer algorithms. Most ICFs are estimated using the plasma simulation code Cloudy \citep[most recently described by][]{ferland2017}, assessing the ionization in one dimension,
assuming a ionizing radiation field incident on a plane-parallel slab of
constant density gas.

Advances in the fidelity of such modeling will address the simplifying assumptions currently being made. The
radiation field is commonly based on a model of the extragalactic ultraviolet
background (UVB) derived from the transfer of radiation from QSOs and star
forming galaxies (with an assumed escape fraction) over redshift
\citep[][]{haardt2012,khaire2019, cafg2019}. Background radiation fields with a
larger contribution from escaping stellar radiation tend to produce lower
abundances \citep[by $\sim0.3-0.5$ dex for several widely-used radiation fields;
][]{fumagalli2016,chen2017,wotta2019}. The contribution from local
radiation sources (e.g., nearby galaxies) can be included, but this typically
requires making poorly-constrained decisions about their relative contribution.
Given these uncertainties, statistically robust constraints on ICFs will likely come
from studies of populations using large samples \citep[e.g.,][]{fumagalli2016,
wotta2019}, folding uncertainties in radiation field into the assessment of
abundance uncertainties. Exploring the ionization of more realistic 3-dimensional geometries, perhaps drawn from hydrodynamic simulations, will also help clarify the uncertainties caused by the simplifying assumptions that are currently made when calculating ICFs.

\subsubsection{Dust Effects: Elemental Depletion}
\label{sec:depletion}

It is well-established that most metals are under-abundant in the interstellar
gas of the Milky Way and that this deficit is a result of the metals'
incorporation into the solid phase, i.e., into interstellar dust. This  {\it
depletion} of metals is differential, \citep{jenkins2009}, with some elements
showing a higher affinity for incorporation into solid-phase grains than others
based on their chemical properties. In total more than 50\% of the metals in the
Milky Way's interstellar medium (ISM) are incorporated into grains. Depletion of
gas-phase metal abundances are clearly also seen in the Large \citep{decia2018a}
and Small \citep{jenkins2017} Magellanic Clouds, though with a smaller degree of
depletion reflecting the lower dust-to-metals mass ratios in these systems.

Given the presence of dust in a broad range of galaxies, depletion is naturally expected to be also detected in the material traced by high-redshift absorption lines. Indeed, some commonly-accessible elements can be subject to large biases:
90-99\% of the Milky Way's interstellar Fe is locked into the solid phase
\citep{jenkins2009}, a bias of $>1$ dex in the gas-phase abundance compared with
the total.  There is strong evidence for the differential depletion of metals in
the neutral absorbers, though typically to a smaller degree than in the Milky
Way \citep[][]{decia2016}. Detections of the 2175 \AA\ absorption ``bump''
\citep{junkkarinen2004} and strong infrared absorption features
\citep{aller2014} associated with solid-phase material at the redshifts of
foreground neutral absorbers provide further support for the presence of dust in
these systems. 

The differential nature of elemental depletion is also an advantage. Early works often used only lightly-depleted elements to derive abundances (e.g., focusing on Zn or to a lesser degree Si). More recent works have taken advantage of
the patterns of differential depletion to correct {\it gas-phase}
measurements to {\it total} metal abundances (e.g., for neutral gas see
\citealt{decia2016, decia2018}; for partially-ionized gas \citealt{quiret2016,
fumagalli2016}).  These efforts follow the spirit of
\citet{jenkins2009}, using empirically-calibrated sequences of relative metal depletions to correct for the metals incorporated into grains \citep{decia2016, jenkins2017, decia2018, decia2018a}.

\subsubsection{Dust Effects: Sample Bias}
\label{sec:samplebias}

Dusty intervening quasar absorbers can prevent the inclusion of the background
quasars from optically-selected samples in the first place. The reddening due to
the dust potentially removes the quasars from color-selected samples or the
extinction causes them to be too faint to be selected. While analysis of
optically-selected quasars provides mixed results on the significance of
reddening by absorbers \citep{frank2010, murphy2016}, there exist examples of
quasars outside of these selection criteria that show reddening by high column
density foreground absorbers \citep{geier2019}.

Quantifying the bias introduced by the optical selection requires alternate
approaches to  identify quasars. Samples of red quasars selected in UKIDSS or
mid-infrared WISE appear to be reddened by material intrinsic to the quasars
themselves \citep[e.g.,][]{maddox2012, krogager2016b}. \citet{ellison2001} were
the first to use a radio-selected quasar sample of $\approx25$ objects to
characterize the differences with optically-selected samples, finding no
significant discrepancies in the absorber statistics. \citet{jorgenson2006}
found a similar result using a larger radio-selected sample; however, they did
identify eight candidate radio quasars (out of their sample of $\sim60$) with no
optical counterparts down to faint limits, hence possible candidates for lines
of sight with dusty absorbers. These samples likely remain too small to put
strong constraints on the presence of high column density, high
metal/dust-content absorbers.

\citet{vladilo2005} use the Galactic extinction law to estimate that while the
dust content of observed absorbers is not high, systems at $z = 1.8-3.0$ with
higher metal and dust columns than found in the current sample may be
responsible for obscuring $\approx30\%$ to 50\% of quasars. In this case, the
highest metal content absorbers are missing from any census, thus significantly
biasing our results. Work by \citet{krogager2019} also including the effect of
color selection argues that current quasar samples possibly underestimate the
mass and metal densities of absorbers by at least 10--50\% and 30--200\%,
respectively. To progress on quantifying this effect directly, promising
alternate paths include quasar selection via luminosity variability
\citep{palanque-delabrouille2016} -- especially in the LSST era, X-ray selected
\citep{merloni2012}, or the absence of proper motions in astrometric surveys
\citep[e.g., with the {\it GAIA} mission;][]{heintz2018a}.

\subsection{Building Statistical Samples of Cosmic Metals}

Samples of high quality quasar spectra key to measuring gas-phase metals have expanded dramatically in the last two decades thanks
in parts to opportunities offered by high-resolution spectrographs on 8--10 m-class telescopes. Publicly available archives of quasars are providing an unprecedented new pool of hundreds of quasar spectra up to
the highest spectral resolution ($R>40,000$). Homogeneous reprocessing of such
datasets provides a goldmine to address new science goals beyond those of the initial observers for both ground-based \citep{zafar2013b, murphy2019, omeara2017}\footnote{UVES/SQUAD: {\tt https://github.com/MTMurphy77/UVES$\_$SQUAD$\_$DR1},\\ HIRES/KODIAQ: {\tt https://koa.ipac.caltech.edu/applications/KODIAQ/kodiaqDocument.html}} and space-based data \citep{peeples2017}.\footnote{HST/COS: {\tt https://archive.stsci.edu/hst/spectral\_legacy/}}
This approach extends to mm wavelengths, with
ALMACAL\footnote{ALMA: {\tt https://www.almacal.wordpress.com}} providing access to a large sample of quasar sight lines \citep{klitsch2020}. We anticipate
that this ultimate part of the data flow, making science-ready data publicly available, will become a key component of observatories in the future.

\subsection{Cosmic Evolution of Neutral Gas Metallicity}
\label{sec:metallicity}

\begin{figure}[h]
\includegraphics[width=5.5in]{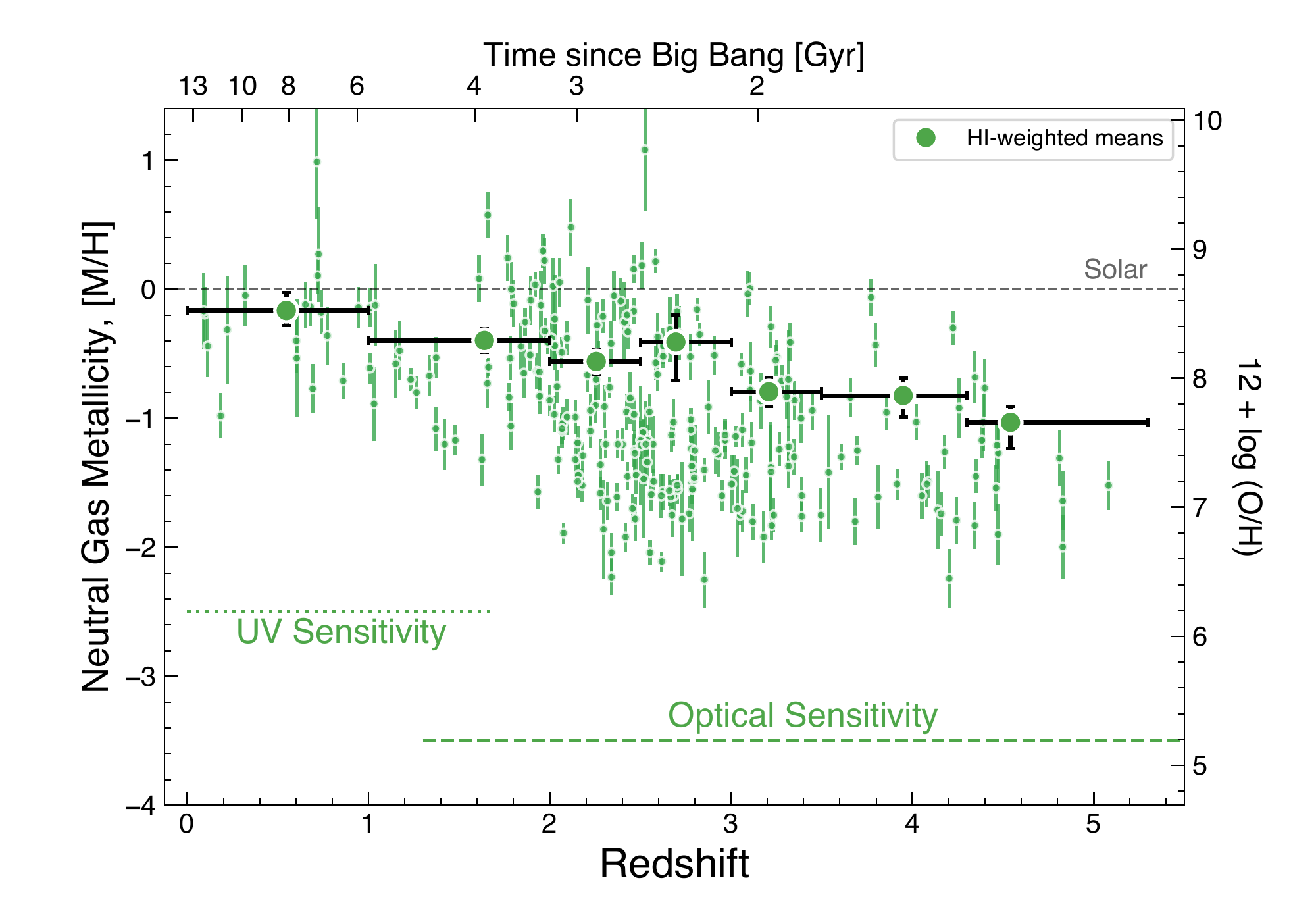}
\caption{Metallicity measurements, expressed as [M/H], for absorbers tracing neutral gas in the Universe ($\logNHI \ge 20.3$). The metallicities include corrections for dust depletion using the empirical approach of \citet{decia2018}. The larger points with error bars are \HI -weighted means with redshift. The increase in the weighted mean metallicity to lower redshift is modest, with a rise of $\sim1$ dex from z $\approx$ 5 until today. The scatter in the population at any redshift is larger than the overall increase in the mean. Lower metallicity gas than that seen in this statistical sample could readily be found: typical sensitivities (shown as dotted/dashed lines) are 1-2 orders of magnitude below the lowest observed metallicities.
}
\label{fig:metallicity}
\end{figure}

We present here a collection of depletion-corrected total metallicity
measurements of neutral gas ($\logNHI \ge 20.3$). This compendium provides a measure of the global mean
metallicity from z $\approx$ 5 to today. Figure \ref{fig:metallicity} shows the metallicities of a select sample of
neutral gas absorbers as a function of redshift. Here we use [M/H] to signify
the generic metallicity. Functionally we adopt ${\rm [M/H]} = {\rm [Si/H]}$ to
tie the metallicity to $\alpha$ elements. Readers should be aware that the non-solar
$\alpha$/Fe ratios in low-metallicity absorbers can make this depart from
[Fe/H]. However, we find that the [M/H] values presented here are close proxies
for the abundances by mass (i.e., $Z_{\rm neutral \ gas} \approx Z_\odot \times
10^{\rm [M/H]}$). Along the right axis we show the corresponding $12+\log ({\rm O/H})$ scale favored in \HII\ region emission line metallicity studies. \HII\ region metallicity determinations probe a different phase of the ISM than the data shown here and are likely affected by self enrichment. Studies comparing neutral and ionized gas metallicities in the same galaxies find they are in good agreement \citep{christensen2014, rahmani2016}.

The majority of the measurements in Figure \ref{fig:metallicity} are drawn from
the dust-corrected sample compiled by \citet{decia2018}.\footnote{Specifically,
we included all systems in Table C2 of \citet{decia2018}. In cases where there
were duplications with Table C2, we favoured the values listed in Table C1. We
added two objects present in Table C1 but not in Table C2 (namely J1009+0731 and
Q0454+039) and removed two systems which are formally below the DLA canonical
definition (J1237+0647 and Q0824+1302). We have complemented the \citeauthor{decia2018} sample with additional measurements at
$z \ga 4$ from the compilations of \citet{berg2016} and \citet{poudel2018} as
well as at $z \la 1$ drawing from \citet{oliveira2014} and the compilation of
\citet{wotta2016}. For these complementary systems, we have derived
dust-corrected metallicities following \citet{decia2018}.}
The individual [M/H] measurements are listed in Supplemental Table 1. The correction for elements locked into dust grains is based on the measurement of the relative amounts of multiple elements, with reference to empirical calibrations using the Milky Way, Magellanic Clouds, and neutral absorbers \citep{decia2016, decia2018}. This approach corrects for the differential elemental depletion into dust, providing the total gas+dust abundance of the neutral absorbers in Figure \ref{fig:metallicity}.
Making corrections for differential depletion requires measurements of more than
one metal.  This sample, through this requirement, leaves off some of the
highest-redshift metallicity measurements,
leading to a  limited number of systems at $z>4.5$.
Figure \ref{fig:metallicity} shows a general trend of increasing metallicity
with decreasing redshift, reaching nearly solar metallicity today. We also note that the scatter of the individual measurements is larger than rise with cosmic time.

We note that the sample presented in Figure \ref{fig:metallicity} is not
sensitivity limited. We show in dashed horizontal lines the limits of
metallicity measurements over two redshift regimes corresponding to the
ultraviolet and optical wavelength windows which are below the lowest data in our compilation.  Our
compendium, which is designed to trace the mean metallicity of the entire
population, intentionally excludes results from specialized searches for
low-metallicity gas (e.g., the $[{\rm M/H}] \sim -3.2$ system identified by
\citealt{cooke2017}). Indeed, beyond using metal tracers, a quest has been
undertaken to identify the chemical signature of the first stars among the most
metal-poor quasar absorbers \citep[see recent work by][and references
therein]{cooke2017}.  Identifying and measuring the detailed chemical abundance
patterns of the most metal-poor absorbers is important to determine the chemical
abundance pattern of the earliest generation of stars and perform a strong and
informative test of nucleosynthesis models of metal-free stars.  Future
prospects of finding extremely metal-poor (and possibly pristine) gas at high redshift are promising with the next generation of  telescopes
\citep{fumagalli2011, cooke2017}.

Figure \ref{fig:metallicity} also shows the \HI -weighted mean metallicity as large dots with error bars. The \HI -weighted mean metallicity provides a robust measure of the global metallicity evolution of the neutral phase of the Universe. We compute column density-weighted mean abundances of $n$ systems in each redshift bin as:
\begin{equation}
	[ \langle {\rm M/H} \rangle ] =
        \log \, \langle {\rm M/H} \rangle -\log \, ({\rm M/H})_\odot
	\label{}
\end{equation}
where
\begin{equation}
	\langle{\rm M/H} \rangle \equiv
        \frac{\sum\limits_{i=1}^{n} 10^{[{\rm M/H}]_i} \NHI_i}
        {\sum\limits_{i=1}^{n} \NHI_i}
	\label{eqn:meanmetal}
\end{equation}
The resulting values are shown in the Figure \ref{fig:metallicity} and
summarized in Table \ref{tab:metallicity}. The horizontal errors show the bin size, while the data points are located at the \HI -weighted mean redshift in each bin. The vertical error bars show the 68\% confidence intervals. Note that these means lie above the
median results for the individual systems both because we calculate the means
from the linear metallicities (Equation \ref{eqn:meanmetal}) and because of the
influence of high column density, high metallicity systems. The binning in
redshift is arbitrary, though with an eye toward including at least $\approx15$
systems in each bin. The \HI -weighted mean metallicity of neutral gas in the Universe increases slowly from high to low redshift, rising by an order of magnitude from z $ \approx $ 5 to 0.

\begin{table}[h]
\begin{center} \tabcolsep7.5pt
\caption{Mean metallicity and dust content of neutral gas absorbers\label{tab:metallicity}}
\begin{tabular}{@{}c|c|c|c|c@{}}
\hline \hline
$\langle z \rangle$ & $z$ range & Num.$^{\rm a}$ & $[\langle {\rm M/H} \rangle]^{\rm b}$ & $\log \langle \dtg \rangle^{\rm c}$ \\
\hline
0.55 & [0.00, 1.00) & 18 & $-0.16^{+0.14}_{-0.11}$ & $-2.57^{+0.17}_{-0.16}$ \\
1.64 & [1.00, 2.00) & 40 & $-0.40^{+0.09}_{-0.09}$ & $-2.87^{+0.11}_{-0.11}$ \\
2.26 & [2.00, 2.50) & 62 & $-0.56^{+0.10}_{-0.11}$ & $-3.02^{+0.13}_{-0.15}$ \\
2.70 & [2.50, 3.00) & 57 & $-0.41^{+0.21}_{-0.30}$ & $-2.75^{+0.25}_{-0.47}$ \\
3.21 & [3.00, 3.50) & 38 & $-0.79^{+0.11}_{-0.12}$ & $-3.32^{+0.15}_{-0.16}$ \\
3.95 & [3.50, 4.30) & 21 & $-0.82^{+0.14}_{-0.17}$ & $-3.43^{+0.17}_{-0.22}$ \\
4.54 & [4.30, 5.30) & 14 & $-1.03^{+0.12}_{-0.20}$ & $-3.62^{+0.14}_{-0.26}$ \\
\hline
\end{tabular}
\end{center} \begin{tabnote} $^{\rm a}$Number of absorbers in each redshift bin; $^{\rm b}$\HI -weighted mean metallicity (\S \ref{sec:metallicity}; Figure \ref{fig:metallicity}); $^{\rm c}$\HI -weighted mean dust-to-gas mass ratios, \dtg\ (\S \ref{sec:dtgdtm}; Figure \ref{fig:dtg_vs_z}).\end{tabnote}
\end{table}

\begin{textbox}[h]
\section{Cosmic Evolution of Neutral Gas Metallicity}
The large samples of high-resolution absorption line measurements collected over the last two decades, coupled with empirical methods to account for the effects of dust grain depletion, provide robust measures of the metallicity up to z $\approx$ 5. We show there is a mild metallicity evolution with cosmic time, but the scatter of the individual measures at a given redshift is larger than the total evolution. The lower bounds of the metallicity distribution are orders of magnitude larger than observational sensitivities at all redshifts: pristine material is exceedingly rare in neutral gas.
\label{box:metallicity} 
\end{textbox}


\subsection{Census of Metals in the Universe}
\label{sec:missing_metal}

Two decades ago, \citet{pettini1999} and other subsequent works noted the
paucity of metals compared with expectations in the available data at
$z\approx2$. Their expectations were based on some of the earliest estimates of
the star formation rate density evolution of the Universe, measurements of metals in neutral gas, and stellar metals from high-redshift galaxies.
This {\it missing metals problem} spurred many follow-up works and thinking about the
global distribution of metals in the Universe. Today we have significantly more
complete information over a broad range of lookback times, covering $\approx90\%$ of the age of the Universe. These include greatly increased information on the evolution of the
star formation rate density \citep[as summarized in ][]{madau2014}, as well as
large samples of quasar absorbers with robust metallicity measurements that
include a consistent treatment of the dust depletion, as summarized in \S
\ref{sec:metallicity}. These facts motivate a reassessment of the cosmic
evolution of the classical missing metals problem.

\subsubsection{Total Metal Mass Density Produced by Stars}
\label{sec:expected_metals}

To compare to the measured metal densities and their evolution, we estimate the total amount of metals produced by star formation as a function of redshift which is developed in the rest of \S \ref{sec:metal}. We scale the stellar mass density in long-lived stars and stellar remnants by an estimated yield following \cite{madau2014}. We write the total comoving metal mass density produced by stars as
\begin{equation}
    \rhomet (z) = y \rhostar (z),
\end{equation}
where $y$ is the integrated yield of the stellar population. In this case the stellar mass density is that remaining in long-lived stars, derived by integrating the star formation rate density, $\psi (z)$, over time / redshift:
\begin{equation}
    \rhostar (z) = (1-R) \int_0^{t(z)} \psi (z) \left| \frac{dz}{dt} \right| dt .
\end{equation}
In this context $R$ is the return fraction, the fraction of the stellar mass that is immediately returned to the gas when massive stars explode. Here we adopt results for a \cite{chabrier2003} initial mass function (IMF), which gives $R = 0.41$ if we assume instantaneous recycling for stars more massive than $m_0 \approx 1 \, \msun$ \citep[][]{madau2014}. We note that the choice of IMF is not affecting the results as the metal production rate is directly related to the mean luminosity density.

We adopt a simplified yield $y = 0.033 \pm 0.010$, drawn from the discussion in
\cite{peeples2014} with an error which encompasses the uncertainties associated
with model assumptions and differing nucleosynthetic inputs.
The Chabrier IMF gives results  consistent with direct observations of stellar
mass density at $z\approx0$  compiled by \citet{madau2014}, though, e.g.,
$\approx0.1 - 0.2$ dex higher than the recent results of \citet{driver2018}.
Adopting a total metal yield ignores complexities of the stellar sites of metal
production (esp., the time lag associated with metals produced by type Ia
supernovae and AGB stars) and the potential for metallicity-dependent yields.
However, we are most interested in the global metal content of the Universe for
times at which several non-primordial populations have contributed metals, and
this yield provides an estimate the {\it total} metal output.

\subsubsection{Metal Mass Densities Traced by Absorption}
\label{sec:omega_metal}

We combine absorption-line derived metal abundances of distant gas with the cosmological density of the same absorbers, $\Omega_{\rm gas}$, to assess the contribution of neutral and more ionised gas to the total metal content of the Universe. Critically, this approach to assessing metal abundances is applied uniformly with redshift, until $z \approx 5$. To calculate the cosmological density of metals associated with absorption line systems, we combine the \HI -weighted mean metallicity of the absorbers with $\Omega_{\rm gas}$:
\begin{equation}
\OmegaMet =  \Omega_{\rm gas} \, \langle Z_{\rm gas} \rangle,
\label{eqn:omegametals}
\end{equation}
where $\langle Z_{\rm gas} \rangle$ is the mean metal abundance by mass in the absorbers. Although we calculate $\langle Z_{\rm gas} \rangle$ explicitly, we find $\langle Z_{\rm gas} \rangle \approx 10^{[\langle{\rm M/H} \rangle_{\rm gas}]} \, Z_\odot$ \citep[e.g., non-solar abundances have a small impact on the calculation; ][]{decia2016}.

\begin{figure}
\includegraphics[width=5.in]{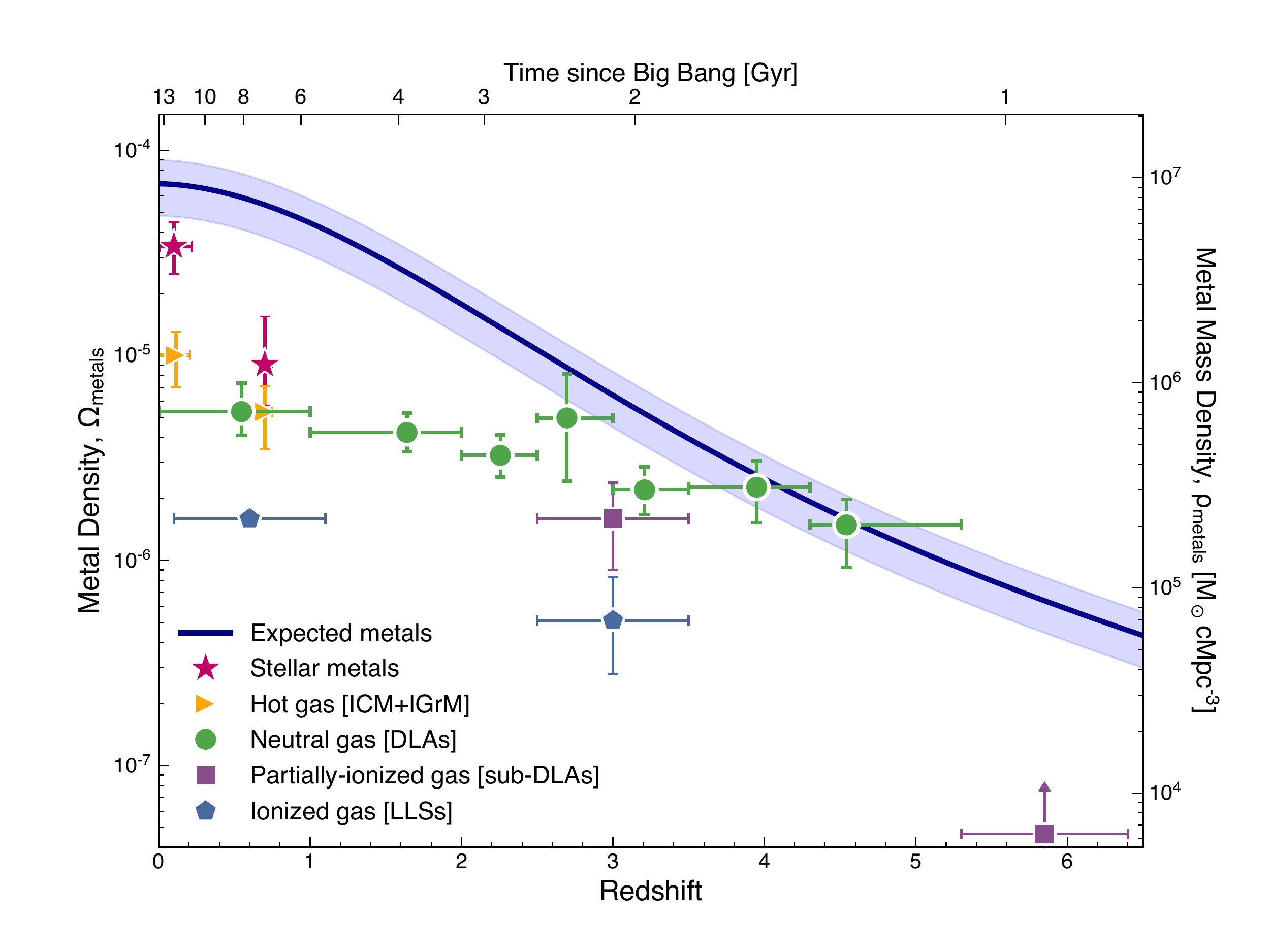}
\caption{Global census of the metals in the Universe traced by the evolution of the cosmic metal densities compared with the expected amount of total metals ever output. The data points focus on robustly-observed contributors. For this reason the data are primarily representative of high density matter and have varying degrees of completeness with redshift. The lower limit at $z\sim6$ is from the \ion{O}{i} measurements of \citet{becker2011}, which we treat as a lower limit to allow for the elements not cataloged. After the dust depletion correction is uniformly applied, we find that the metallicity in the neutral gas dominates the metallicity of ionised gas.
}
\label{fig:omega_metal}
\end{figure}

Figure \ref{fig:omega_metal} shows a non-exhaustive sample of the cataloged
contributions to the cosmological metal density as a function of redshift.
The metal density associated with neutral gas absorbers (the absorbers with
$\logNHI \ge 20.3$) is shown with the same redshift bins as Figure
\ref{fig:metallicity}. These values represent the total metals in these systems,
i.e., including both the gas and dust (see also \S \ref{sec:dust} below). These
metal densities associated with neutral gas are summarized in Table
\ref{tab:omega}. Contributions from lower column density absorbers are taken
directly from their respective references, from \citet{fumagalli2016} for the $z
\approx 3$ results and \citet{lehner2019} for $z \la 1$. At the highest
redshifts, probing the metals in absorption becomes difficult as the flux
bluewards of the quasar \lya\ emission is suppressed by \laf\ absorption.
\citet{becker2011} measure the density of \ion{O}{i} absorption at
$z\approx6$; we plot this point assuming $\OmegaMet \ga \Omega_{\rm OI}$ to
allow for the elements not cataloged (O is $\approx40\%$ of the solar metal
mass). This point traces gas with $\logNHI \ga 19.0$, and likely is dominated by
the lower column density end of this range; thus we associate it with the
partially-ionized absorber range.

\begin{marginnote}[]
    \entry{ICM}{Intracluster Medium}
    \entry{IGrM}{Intragroup Medium}
\end{marginnote}

At lower redshifts where stars and hot gas found in groups and clusters play an
important role, we also include the density of metals captured in stars and
in the hot gas associated with clusters and groups of galaxies. We derive the
quantity of metals sequestered in stars using the stellar metallicities as a
function of galaxy stellar mass measured by \cite{gallazzi2008} and
\cite{gallazzi2014} at $z \sim 0$ and $z \approx 0.7$, respectively, using the
galaxy stellar mass function from \citet{wright2018} (errors from Monte Carlo
sampling). We derive the combined intragroup medium (IGrM) and intracluster
medium (ICM) metal densities using a mass integral of the product of the
halo mass function \citep[][as implemented in the COLOSSUS code of
\citealt{diemer2018a}]{bocquet2016}, the hot gas baryonic fraction with mass
\citep{chiu2018}, and a global metallicity $Z_{\rm IGrM} \approx Z_{\rm ICM}
\approx \tfrac{1}{3} Z_\odot$, assumed to be constant with redshift and halo
mass \citep{mantz2017, yates2017}. For both the stellar and hot gas (ICM+IGrM)
cases we assume our adopted solar metal abundance.

%

\begin{table}[h]
\begin{center} \tabcolsep7.5pt
\caption{Cosmic mass densities in neutral absorbers \label{tab:omega}}
\begin{tabular}{@{}c|c|c|c|c|c|c@{}}
\hline \hline
$\langle z \rangle$ & $z$ range & Num.$^{\rm a}$ & $\log \OmegaGas$$^{\rm b}$ & $\log \OmegaMet$$^{\rm c}$ & $\log \OmegaDust$$^{\rm d}$ & $\log \OmegaMet^{\rm expected}$$\ ^{\rm e}$ \\
\hline
0.55 & [0.00, 1.00) & 18 & $-3.23^{+0.01}_{-0.01}$ & $-5.27^{+0.14}_{-0.12}$ & $-5.79^{+0.17}_{-0.16}$ & $-4.24\pm0.13$ \\
1.64 & [1.00, 2.00) & 40 & $-3.10^{+0.01}_{-0.01}$ & $-5.38^{+0.09}_{-0.09}$ & $-5.97^{+0.11}_{-0.11}$ & $-4.59\pm0.13$ \\
2.26 & [2.00, 2.50) & 62 & $-3.04^{+0.01}_{-0.01}$ & $-5.49^{+0.10}_{-0.11}$ & $-6.07^{+0.13}_{-0.15}$ & $-4.86\pm0.13$ \\
2.70 & [2.50, 3.00) & 57 & $-3.01^{+0.01}_{-0.01}$ & $-5.30^{+0.21}_{-0.31}$ & $-5.77^{+0.25}_{-0.47}$ & $-5.06\pm0.13$ \\
3.21 & [3.00, 3.50) & 38 & $-2.98^{+0.02}_{-0.02}$ & $-5.66^{+0.11}_{-0.12}$ & $-6.30^{+0.15}_{-0.15}$ & $-5.29\pm0.13$ \\
3.95 & [3.50, 4.30) & 21 & $-2.94^{+0.02}_{-0.02}$ & $-5.64^{+0.13}_{-0.17}$ & $-6.37^{+0.17}_{-0.23}$ & $-5.58\pm0.13$ \\
4.54 & [4.30, 5.30) & 14 & $-2.91^{+0.02}_{-0.02}$ & $-5.83^{+0.13}_{-0.21}$ & $-6.53^{+0.15}_{-0.27}$ & $-5.79\pm0.13$ \\
\hline
\end{tabular}
\end{center} \begin{tabnote} $^{\rm a}$Number of absorbers in each redshift bin; $^{\rm b}$Cosmic gas mass density at $\langle z \rangle$ (\S \ref{sec:omega_neutral}; Figure~\ref{fig:omega_neutral}); $^{\rm c}$Cosmic metal mass density in neutral gas (\S \ref{sec:omega_metal}; Figure \ref{fig:omega_metal}); $^{\rm d}$Cosmic dust mass density in neutral gas (\S \ref{sec:omega_dust}; Figure \ref{fig:omega_dust}); $^{\rm e}$Total expected cosmic metal mass density (\S \ref{sec:expected_metals}; Figure \ref{fig:omega_metal}). \end{tabnote}
\end{table}

Figure \ref{fig:omega_metal} therefore provides a global census of metal content
of the Universe over a large look-back time. It includes only values that are
the most robustly and directly measured. Thus, there are some redshift ranges
for which fewer contributions are included and some contributions are not
represented at all. The expected metals follow the stellar density of the
Universe, rising by $\approx2$ orders of magnitude between $z \approx 5$ and
$z\approx 0$.  Consistent with the greatly diminished star formation rate
density at low redshift, the curve flattens significantly at $z<1$. By contrast,
we see the metal density associated with neutral gas shows a much shallower
rise with time, increasing by only a factor of $\approx8$ since $z\approx 5$.
Remarkably, the metals associated with neutral gas alone are consistent with all
the expected metal density of the Universe at $z \ga 2.5$
\citep{rafelski2014}. We note in passing that this leaves little room for a
large population of metal-rich absorbers missed due to dust (\S
\ref{sec:samplebias}).

\begin{figure}
\includegraphics[width=5.in]{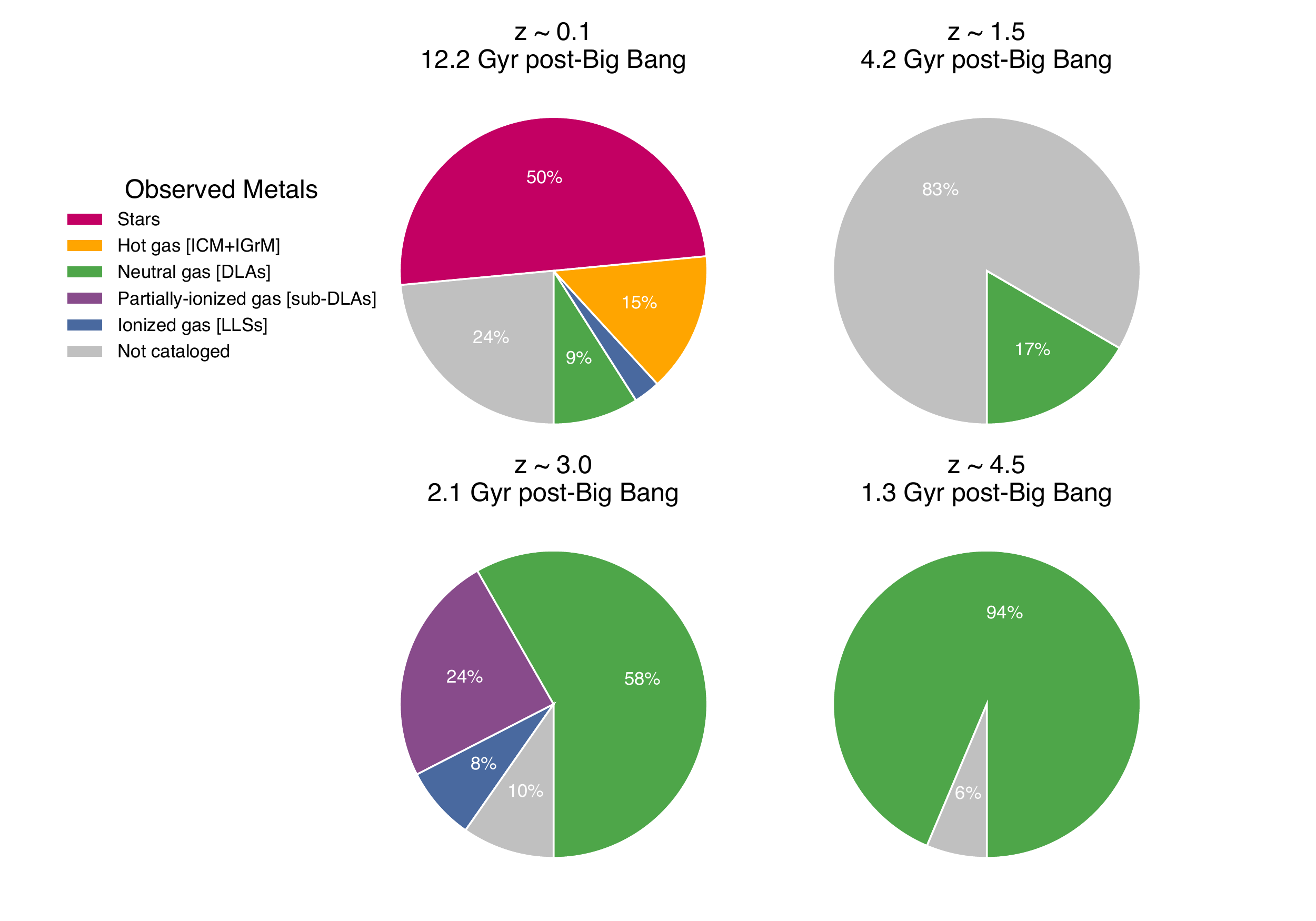}
\caption{A modern census of metals showing the fractional contribution of
directly-measured metal densities to the total expected based on the
stellar densities at four redshifts. The estimate of the total metal mass
densities assumes a yield $y =0.033\pm0.010$; thus, when the proportion of the
expected metals not cataloged in our census (grey) is $\la30\%$, the cataloged
metals are consistent with expectations.
At redshifts $z\ga 2.5$, the vast majority of the expected metals are found in
low-ionization gas in all its forms. At lower redshift, there is a greater
diversity of contributors, though nearly all of the metals have been cataloged.
At intermediate redshifts ($1 \la z \la 2$) the likely contributors are not yet
robustly determined. Overall, the expected metal content of the Universe is largely accounted for.}
\label{fig:omega_metal_pie}
\end{figure}

This is further demonstrated in Figure \ref{fig:omega_metal_pie}, which shows
the fractional contributions of the cataloged components to the expected metals
at four distinct redshifts, $z = 0.1, 1.5, 3.0,$ and 4.5. At the higher
redshifts, in the range $2.5<z<5$, essentially all of the expected metals are
found in the cataloged classes of absorbers, including the partially-ionized
($19.0 \le \logNHI \le 20.3$) and ionized absorbers ($17.2 \le \logNHI \le
19.0$).  Some previous reports found that the partially-ionized gas contributes
more to \OmegaMet\ than the neutral gas, whereas Figure \ref{fig:omega_metal}
shows the opposite. The differences arise from different treatments of the
inclusion of dust corrections in metallicity estimates. Overall, when dust
corrections are applied to both column density ranges, all works agree the
contribution of partially-ionized gas is less than the neutral gas \citep[][this
work]{peroux2007, kulkarni2007}.

In general, the contributors to metals are more diverse at lower redshifts: stars, the hot gas (ICM+IGrM), and the ionised phase of the gas all become more important. The metals in the neutral phase contribute only a small fraction of the expected metals. Figure \ref{fig:omega_metal_pie} shows that at $z\approx0.1$, half of the metals are in stars and the metals in the hot gas (ICM+IGrM) become comparable to the metals in the neutral gas at $z \la 1$.

\subsubsection{Other Sinks of Metals}
\label{sec:missing_census}

Figures \ref{fig:omega_metal} and \ref{fig:omega_metal_pie} show that the metal
densities cataloged here fall short of expectations at $z\la 2.5$. We have
focused on the components with the most robust measurements of their associated
\OmegaMet. However, there is existing guidance on components we have not
included in our census. Working to progressively higher overdensities, the
additional contributors include: the diffuse, photoionized intergalactic medium
\citep[the \laf,][]{mcquinn2016} and the low
density and hot circumgalactic matter \citep[CGM,][]{tumlinson2017}. We note that the warm-hot intergalactic medium (WHIM) in simulations was originally identified as a large baryon reservoir, including 30-50\% of the baryons at low redshift \citep{cen1999, dave2001}. In today's view, most of the metals originally connected to the WHIM are identified with the CGM \citep{werk2014}.

At the lowest density scale, the \laf\ traces diffuse intergalactic gas in the
dark matter filaments threading the Universe (at \HI\ column densities well
below $\logNHI \le 17.2$). At high-redshift, the \laf\ houses the majority of
the baryons in the Universe and is photoionized material (with $T \sim 10^4$ K).
However, the metal content of this matter is low. For example, the results of
\citet{dodorico2013} imply $\OmegaMet^{\rm Ly\alpha} \approx 1.4\times10^{-7}$
at $z\approx 3$ (where we have corrected their estimate from carbon to the total
metal content). This contribution is unimportant compared with the neutral gas
at these redshifts. At low-redshift, diffuse large-scale structures house both
photoionized and collisionally-ionized baryons.  The metal density of these
constituents is assessed through censuses  of metal lines associated with low
column density \HI\ absorbers. At $z\la 1.5$, \citet{shull2014} argue for
$\OmegaMet^{\rm Ly\alpha} = (1.1\pm0.6)\times10^{-5}$.

The denser gas on the scale of galaxy halos also contains  highly-ionized
material that houses significant quantities of metals \citep{tumlinson2017}.
This gas, traced by \siiv, \civ, \nv, and \ovi, likely arises through
photoionization of low density gas as well as in interfaces of isolated cold gas
(clouds) with a hotter medium, in shocks propagating through these clouds, or in
diffuse $T \ga 10^5$ K gas \citep[e.g.,][]{lehner2014, stern2016}. The
uncertainties in the ionization mechanisms and ion fractions limit the precision
with which the total metal budget of this material is estimated.  For
predominantly neutral gas ($\logNHI \ga 20.3$), the contribution of associated
high ionization species to their metal budget is of order 10\%
\citep[][]{lehner2014}. The fraction of metals associated with these
high-ionization species increases at lower \HI\ column densities. For the
partially-ionized and ionized gas, these metals likely become dominant. At
$z\approx2.5$--3.5, \citet{lehner2014} estimated $\OmegaMet \approx (1.5 -
6)\times10^{-6}$.

\subsubsection{Is There a Missing Metals Problem?}

The ``missing'' metals problem poses the question: which baryonic components of
the Universe host the metals? In the results summarized in Figures
\ref{fig:omega_metal} and \ref{fig:omega_metal_pie}, we are observing the
greatest sinks of metals change. At high redshift, $z\ga2.5$, the neutral gas
houses nearly all of the available metals, while at low redshift, $z<1$, the
metals are distributed nearly equally between stars and a broad range of gaseous
environments. This reflects the build up in the typical mass scale of halos
toward low redshift and the influence of energetic feedback over time.

These results are distinguished from historical measures
\citep[e.g.,][]{pettini1999,pagel1999,ferrara2005} in several ways. First, the
results presented here span the full range of redshifts $0 < z \la 5$,
covering $\approx 90\% $ of the age of the age of the Universe. At higher
redshifts ($z\ga2.5$), the metals associated with high column density absorbers
are readily able to match the expected quantity of metals made by stars. The
historical results were constrained to a redshift range at which the expected
metals start to diverge significantly from the metal densities in neutral gas
\citep[$z \approx 2$; ][]{pettini1999, ferrara2005, bouche2007}. At low redshift
($z\la 1$), it is likely that metals found in stars, the hot gas in groups and
clusters, highly-ionized CGM gas, and the \laf\ each contribute metals in
quantities similar to that seen in the neutral gas, which is sufficient to then
close the budget over this redshift regime.

There are clearly redshift ranges for which a full, robust accounting is not
available (e.g., the $z \sim 1.5$ panel of Figure \ref{fig:omega_metal_pie}) due
to the difficulty in observing the relevant metals in this regime. Using the
lower redshifts as a guide, we expect the metal budget to be closed through
deeper NIR spectral surveys of stellar metals, deep X-ray observations of hot
gas in the IGM+IGrM, and deep UV surveys of CGM and \laf\ metals to
higher redshift.
Probing the gas to higher redshifts will require new facilities.
The next generation of X-ray facilities ({\it Athena, Lynx}) will provide robust
measurements of \OmegaMet\ in the ICM up to $z \approx 2$ \citep{cucchetti2018} and allow some absorption line measurements of the ionized phase. Proposed UV facilities ({\it LUVOIR, HabEx}) will allow constraints on IGM and CGM metals to intermediate redshifts.
Between the contributions summarized in Figures \ref{fig:omega_metal} and
\ref{fig:omega_metal_pie} and the additional tracers of lower-density and/or
hotter matter in \S \ref{sec:missing_census}, the expected metal content of the
Universe is largely accounted for.

\begin{textbox}[h*]
\section{Update on the Missing Metals Problem}
\label{box:omega_metal} 
We have compiled a census of metals in in the Universe and compared this to a new assessment of the total metal production by star formation \citep{madau2014}. At high redshift, $z\ga2.5$, neutral gas in the Universe contains most of the expected metals. At low redshift, $z\la 1$, the set of contributors is more diverse, and stars are the dominant contributor. At intermediate redshifts, the likely contributors are yet to be fully characterized; upcoming efforts will account for some of these (e.g.,  stars), while a complete census will require new facilities, notably in the UV and X-ray. Overall, the expected metal content of the Universe is likely accounted for, in contrast to the missing metals problem identified 20 years ago \citep{pettini1999, pagel1999}.
\end{textbox}

\section{COSMIC EVOLUTION OF DUST}
\label{sec:dust}

A substantial fraction of all the metals produced in the Universe are locked
into solid-phase dust grains. Such dust not only has a strong influence on the
observational properties of galaxies, but it is critical in the thermal balance
of gas as well as in shielding the cores of dense clouds from UV radiation,
allowing the formation of molecules critical to the star formation process.
Hence the cosmic evolution of dust mass is a fundamental measure of galaxy
evolution. There are still significant unknowns in the manner in which dust is
created and ultimately destroyed. Thus significant effort has been put into
understanding the galactic scaling relationships that apply to dust, including
the relationship between local metallicity and dust content -- traced through
the dust-to-gas ratio (\dtg) or dust-to-metals ratio (\dtm), as well as the
global dust content of the Universe traced through the dust density,
\OmegaDust. In this section, we use the corrections for dust depletion from \S
\ref{sec:metal} to estimate the dust properties of neutral absorbers and
estimate the dust density evolution of the Universe.

\begin{marginnote}[]
    \entry{\dtg}{\\dust-to-gas ratio}
    \entry{\dtm}{dust-to-metals ratio}
\end{marginnote}

\subsection{Basic Principles of Assessing the Dust in the Universe}
\label{sec:dtgdtm}

The characterization of the bulk statistical properties of dust involve
assessing the \dtg\ or \dtm. The former is the fraction of the interstellar mass
locked into dust grains; the latter is the fraction of the
metal mass incorporated into the solid phase. In the Milky Way, the global
values of these quantities -- based on modeling the infrared emission and
optical/UV extinction -- are $\dtg \approx 0.006$ and $\dtm = \dtg\ Z_\odot^{-1}
\approx 0.45$ \citep[e.g., ][]{draine2007}. Significant work has been put into
characterizing these quantities in galaxies beyond the Milky Way, assessing
observational constraints both on the integrated values
\citep[e.g.,][]{remy2014, de-vis2019} and spatially-resolved values within
galaxies \citep[e.g., ][]{vilchez2019}. These works have placed particular
emphasis on the variation in \dtg\ and \dtm\ ratios with metallicity, stellar
mass, star formation rate, and gas content of the galactic environments, as
these help shape our understanding of the factors that drive
the formation/destruction balance of dust.

The dust depletion corrections used to assess the total metal content of neutral absorbers in \S \ref{sec:metal} have an important by-product: the derived corrections are measures of the metal content of dust grains in the environments traced by
the absorption lines. Parallel to our derivation of $\OmegaMet$, we use these
results to assess the cosmological dust density of the Universe:
\begin{equation}
\OmegaDust \equiv \rho_{\rm dust} / \rhocrit =  \langle \dtg \rangle \, \Omega_{\rm gas},
\label{eqn:omegadust}
\end{equation}
where $\langle \dtg \rangle$ is the mean dust-to-gas mass ratio. The \dtg\
ratios for individual absorption systems are derived from elemental depletions;
because the behavior of each metal species varies \citep{jenkins2009,
jenkins2017}, we build up the total \dtg\ on an element-by-element basis. This
starts from an estimate of the depletion, \deltaX, for each individual element,
$X$, derived from the dust sequences of \citet{decia2018}. The (logarithmic)
depletion is defined:
\begin{equation}
    \deltaX = \log N(X)/N({\rm H}) - \log Z^{X}_{{\rm abs}} / Z^{X}_\odot,
\end{equation}
where $Z^{X}_{{\rm abs}}$ is the abundance by mass of $X$ in the absorber and $Z^{X}_{\odot}$ the equivalent in the Sun. The dust-to-metal ratio for an individual element, $X$, is related to its depletion:
\begin{equation}
    \dtm_X = 1 - 10^{\deltaX}
\end{equation}
\citep[e.g.,][]{vladilo2004,decia2016}. The dust-to-gas mass ratio for an individual element can then be written as
\begin{equation}
    \dtg_X = \dtm_X \, Z^{X}_{{\rm abs}},
\end{equation}
which is related to the measured [M/H] through
\begin{equation}
    \dtg_X =  (1 - 10^{\deltaX}) \, Z^{X}_{{\rm abs}} =
           (1 - 10^{\deltaX}) \, (10^{[{\rm M/H}]} \, 10^{[X/{\rm M}]} Z^{X}_\odot).
\end{equation}
Here we use the intrinsic metallicity of the absorber, [M/H], and allow for
non-solar relative abundances $[X/{\rm M}]$. Ultimately this provides the mass
of an element $X$ incorporated in dust relative to the total gas mass in the
absorber. The total \dtg\ is the sum over all elements, $\dtg_{\rm abs} = \Sigma
\, \dtg_X$.

Practically speaking, it is sufficient to consider only $X = \{ {\rm
C, O, Si, Mg, Fe} \}$ in the sum, as together these represent the vast majority
of the dust mass \citep[see][]{draine2011}.
The depletions of each of these five elements are coupled through the dust
sequences derived by \cite{decia2018}. Observations of relative metal
abundances constrain the overall level of depletion, characterized using
$\deltaFe$ as reference. The observed metals then predict the depletions of all
of the metals. We assume a solar mixture of elements for all but Fe, which shows an empirical under-abundance compared with the $\alpha$ elements \citep{decia2016}, and
[C/Fe]$\, \approx 0$, allowing for an under-abundance of C in lower-metallicity
systems \citep[e.g.,][]{jenkins2017}.

Carbon makes up $\sim30\%$ of the dust mass in the Milky Way
\citep{draine2011}, but there are not good measurements of C/H in quasar absorbers from which to build up a depletion sequence, as lines from the dominant ionization states of C are typically saturated or too weak to observe. Observations of C absorption in the Milky Way are also limited, for the same reasons. Even though these have significant uncertainties, we adopt the relationship between $\deltaFe$ and $\delta_{\rm C}$ from \cite{jenkins2009}.

\begin{figure}[ht]
\includegraphics[width=3.5in]{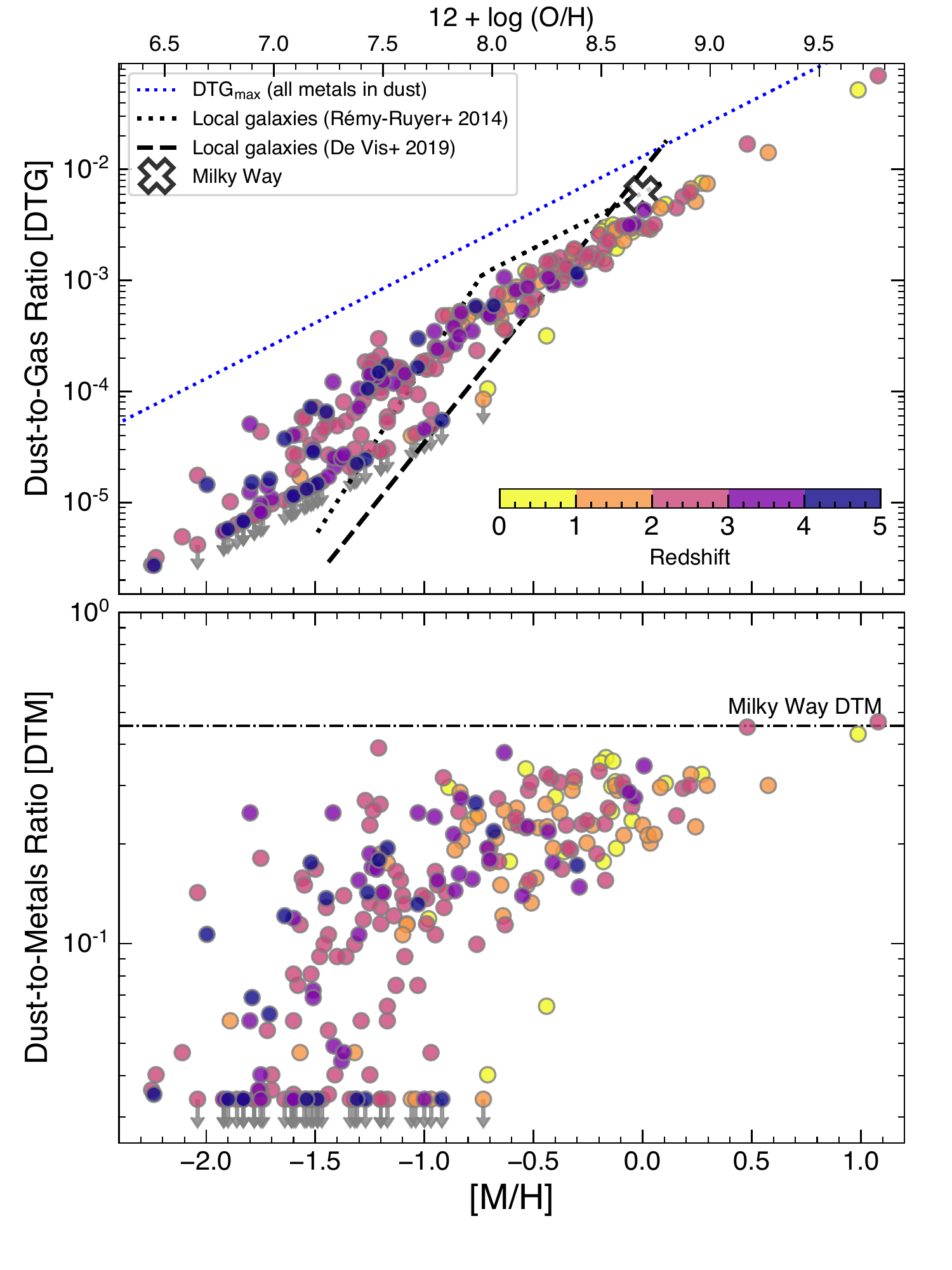}
\caption{Properties of dust in neutral gas as a function
of absorber metallicity. {\it Upper:} The dust-to-gas mass ratios, \dtg, color coded by redshift. The dotted line is the limit in which all metals are incorporated into grains ($\dtm = 1$). The position of the Milky Way is shown. The black lines show fits to the properties of local galaxies derived from far-infrared emission \citep{remy2014, de-vis2019}. The \dtg\ values derived for neutral gas over all redshifts follow trends seen in these low-redshift galaxies, extending these measurements to a lower metallicity regime.   {\it Lower:} The dust-to-metal mass ratio, \dtm, giving the fraction of all of the metal mass bound
into dust. The dot-dashed line represents the \dtm\ ratio in the
Milky Way \citep{draine2007, draine2014}. Upper limits are due to the uncertainty in carbon
depletion. With decreasing metallicity, the \dtm\ values both decrease and show increased scatter, reflecting complex dust chemistry at work in these low-metallicity environments. }
\label{fig:dtgdtm}
\end{figure}

The characteristics of the \dtg\ and \dtm\ ratios as a function of absorber metallicity are summarized in Figure \ref{fig:dtgdtm} and are listed in Supplemental Table 1. The upper panel shows the \dtg\ ratio with metallicity. The dotted line is the maximum allowed \dtg\ ratio, where all the metals are incorporated into grains. The position of the Milky Way is shown by the large cross. Also shown are best-fit trends in \dtg\ ratios for local galaxies derived from far-infrared observations by \citet{remy2014} and \citet{de-vis2019}; while there are many such results, these are typical of the range of results seen at low redshift. The bottom panel shows the \dtm\ ratio as a function of metallicity. Where carbon is the only significant contributor to the dust content, we treat these values as upper limits.

The \dtg\ ratio is a strong function of metallicity, as expected. It is noteworthy that the neutral gas follows the trends seen in low-redshift galaxies at high metallicities ($\MtoH \ga -1$); there is little evidence that the \dtg\ behaves differently with redshift. The neutral gas measurements extend to a metallicity regime lower than found in low-redshift galaxies. A portion of the population shows \dtg\ ratios that extend the trend seen at higher metallicity, while a significant fraction falls well below this trend due to lower \dtm\ ratios.  This suggests a change in the dust assembly at low metallicities.

\begin{figure}[ht]
\includegraphics[width=4.in]{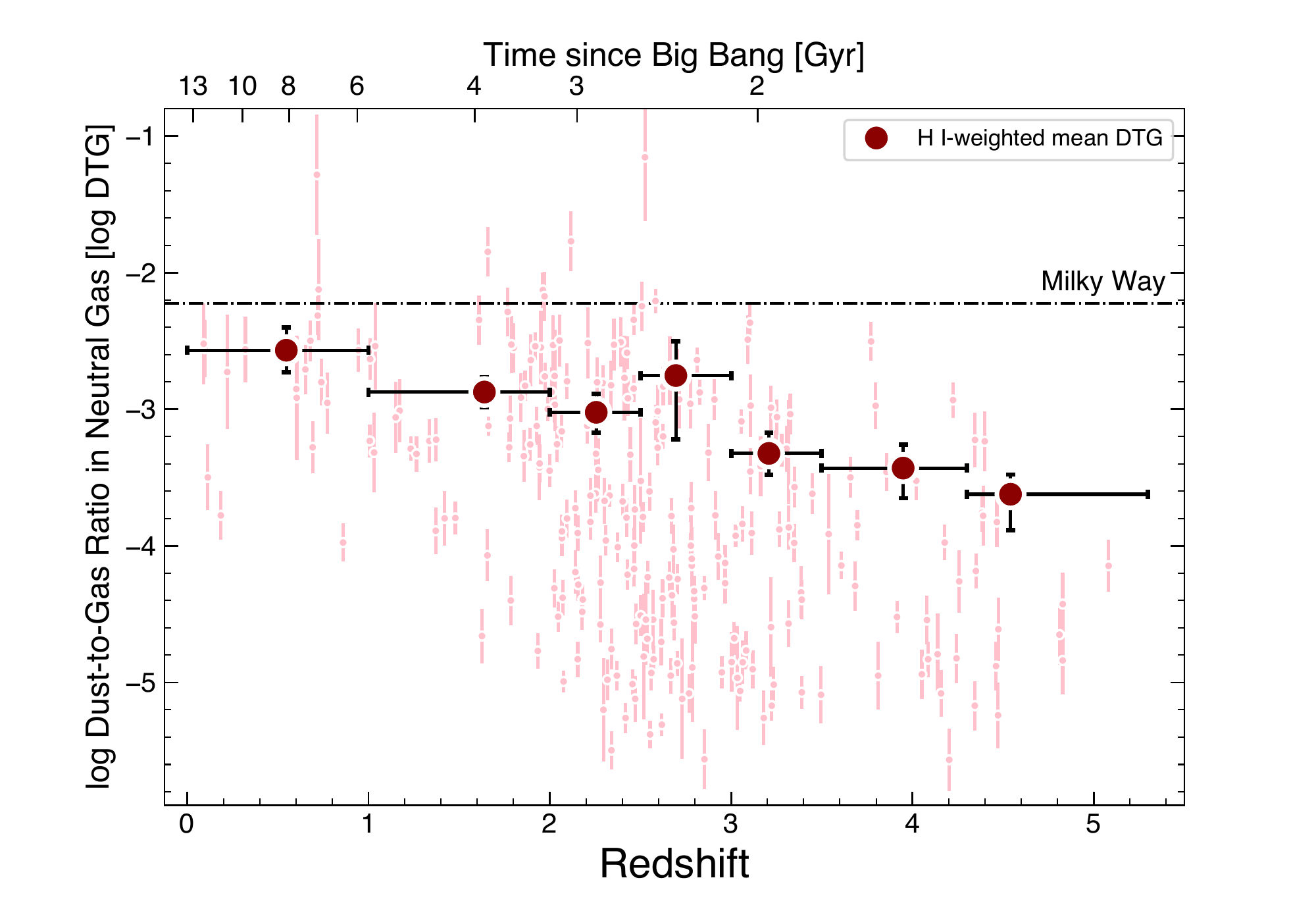}
\caption{The dust-to-gas mass ratio, \dtg, in neutral gas absorbers as a function of redshift. The mean values weighted by \HI\ column are shown as large points with error bars. The Milky Way value from \citet{draine2007} renormalized following \citet{draine2014} is shown. The evolution in redshift mirrors that of the metals that provide the source material for grains (see Figure \ref{fig:metallicity}), with the mean \dtg\ increasing by $\sim1$ dex from z $\approx$ 5 to z=0. The \dtg\ values show an even larger spread at a given redshift than the metallicities. This reflects the large scatter in the \dtm\ at low metallicities and the broad range of metallicities probed by these absorption line measurements.
}
\label{fig:dtg_vs_z}
\end{figure}

We plot the \dtg\ measurements for individual absorbers in Figure
\ref{fig:dtg_vs_z} as function of redshift, in parallel to Figure
\ref{fig:metallicity} summarizing the metallicity evolution of these systems.
The \HI -weighted mean values are shown as larger points with error bars.
The characteristic \dtg\ for the Milky Way is also shown, $\dtg_{\rm MW} \approx
0.45 Z_\odot$ \citep[][renormalized following recommendations in \citealt{draine2014}]{draine2007}. The mean \dtg\ ratio increases by $\sim1$ dex from z $\approx$ 5 until today, a result of the increase in mean metallicity.

\subsection{Cosmological Dust Mass Densities}
\label{sec:omega_dust}

\begin{figure}[h]
\includegraphics[width=5in]{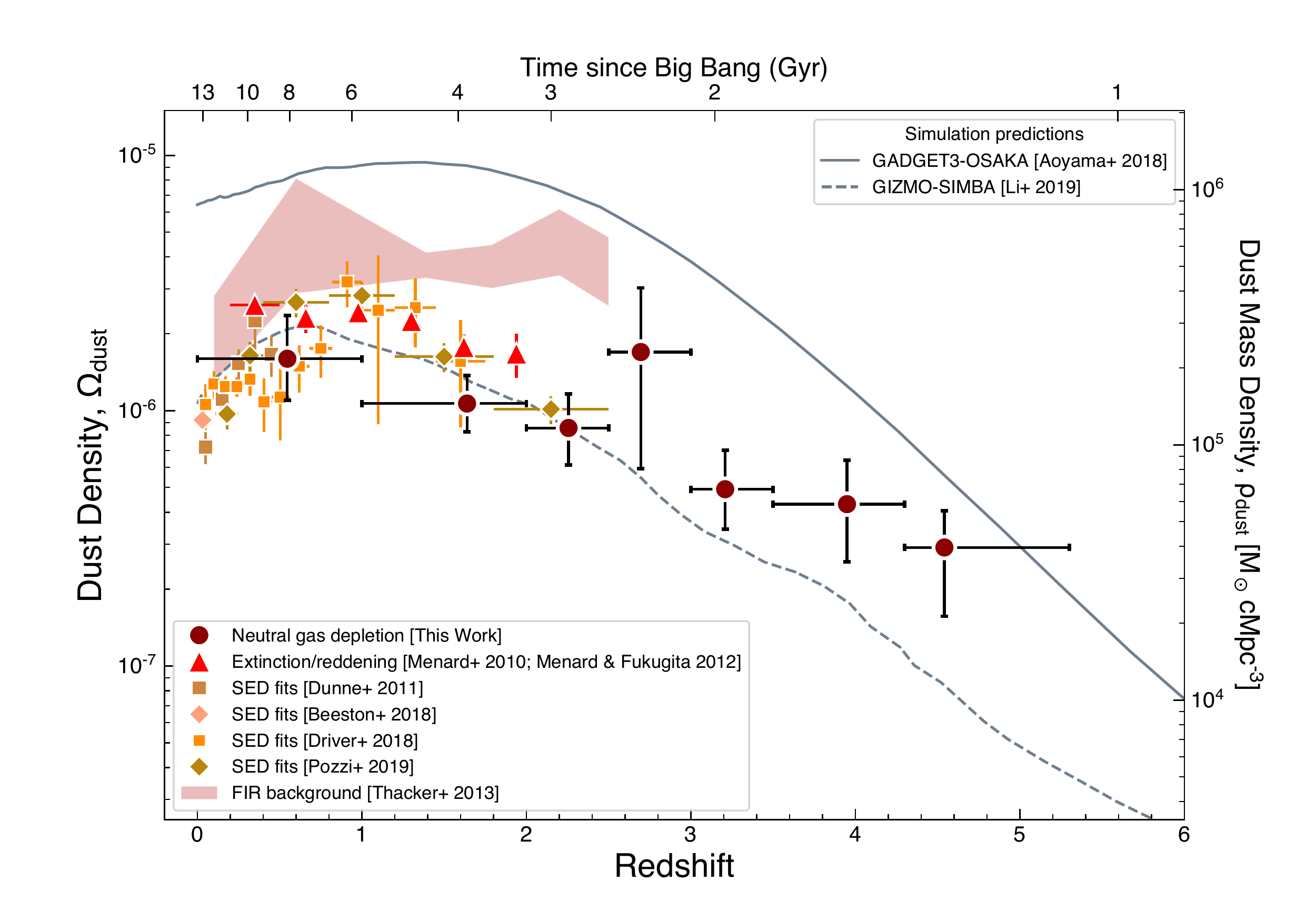}
\caption{Evolution in cosmological dust density, \OmegaDust, derived through
different approaches. Our assessment of the cosmological dust mass density for
neutral gas is shown with large points with error bars. We show several measures
of \OmegaDust\ derived from large samples of individual galaxies; their dust
content is assessed using attenuations derived from optical SED fits or from  SED fits to optical through FIR emission. Estimates of the dust mass density
from the power spectrum of the far-infrared background are shown as the shaded
band. Dust in and around galaxies (esp. in the CGM) is probed through the
reddening or extinction associated with \mgii\ systems or foreground galaxies.
Recent predictions from cosmological hydrodynamic simulations incorporating
on-the-fly dust tracking are also shown by lines. The cosmological dust mass
density associated with neutral gas (which is inclusive of neutral gas in the
ISM and CGM of galaxies) increases by nearly a dex from $z\approx5$ to $z\approx1$,
following a trend similar to that for \OmegaMet\ in neutral gas. At $z<2$, the
measurements from disparate techniques show good agreement.  Measurements of
dust in neutral gas give a uniform observational constraint on the global
build-up of dust from $z\approx5$ to today.}
\label{fig:omega_dust}
\end{figure}

The global evolution of the dust mass in the Universe is captured by the dust density, \OmegaDust. By analogy to the metal densities, this is the comoving density of dust in the Universe normalized by the critical density, $\OmegaDust \equiv \rho_{\rm dust} / \rhocrit$. We derive values of \OmegaDust\ in the neutral gas of the Universe following Equation \ref{eqn:omegadust}. We adopt \HI -weighted mean \dtg\ ratios, which we write in parallel to the mean metallicity of Equation~\ref{eqn:omegametals}:
\begin{equation}
    \langle \dtg \rangle = \frac{\Sigma(\dtg\ \times \NHI)}{\Sigma \NHI}
\end{equation}
Figure \ref{fig:omega_dust} shows the values of \OmegaDust\ in neutral gas derived in this way from $z \approx 5$ to today. Systems possibly missed due to the dust sample bias caveat mentioned in \S
\ref{sec:samplebias} are not accounted for. The evolution of \OmegaDust\ follows
a similar trend as that for \OmegaMet\ in neutral gas, increasing by a factor of
$\approx7$ between the z $\approx$ 5 and today.
That the dust tracks the evolution of the metal density is not surprising,
as the dust mass in any individual system is related to its metallicity (e.g.,
Figure \ref{fig:dtgdtm}). The overall shape resembles estimates of \OmegaMol\ (Figure~\ref{fig:omega_all}).

Figure \ref{fig:omega_dust} also shows \OmegaDust\ derived through other means,
of which there are three main approaches. The first approach integrates dust
mass functions derived from modeling the spectral energy distributions (SEDs) of
individual galaxies as a function of redshift. Selected results derived in this
way are shown in Figure \ref{fig:omega_dust} \citep{dunne2011, beeston2018,
driver2018, pozzi2019}. They rely on the inclusion of mid- or far-infrared
measurements from, e.g., {\it Spitzer}, {\it WISE}, and {\it Herschel}, as well
as detailed models to assess the total dust mass of galaxies.  The results from the
three studies in Figure \ref{fig:omega_dust} are in respectable agreement over
their common redshift range. The second approach derives and integrates over a
luminosity function of galaxies from the power spectrum of the far-infrared
background radiation \citep{de-bernardis2012, thacker2013}. Results derived from
{\it Herschel} observations of the background are shown as a shaded region in
Figure \ref{fig:omega_dust}. The third approach assesses the density of dust
based on its reddening or extinction of background optical sources (triangles in
Figure \ref{fig:omega_dust}).  \cite{menard2010} derived the dust profiles about
$z\sim0.3$ galaxies, while \cite{menard2012} derived \OmegaDust\ using the
extinction due to  \mgii\ absorbers. A fraction of the \mgii\
absorbers selected by \citet{menard2012}, with equivalent widths EW$>$0.8 \AA,
trace the neutral gas studied here ($\logNHI \ge 20.3$;
\citealt{rao2017}). The remainder is associated with lower-density
circumgalactic gas. We see good agreement over the redshift range of overlap in
the values of \OmegaDust\ derived for neutral gas with those derived through
other means.  However, the depletion-based results reported here trace \OmegaDust\ to
significantly higher redshifts than the other methods, providing a measure of
the build-up of dust from z $\approx$ 5. The
depletion-based results for neutral gas are based on counting atoms incorporated
into grains and make no assumptions on the physical properties (such as opacity
or temperature) of the dust.

Taking the whole of the observations, \OmegaDust\ peak around $z\sim1$ ($\sim$8 Gyr ago), suggesting that dust
formation is either concurrent with star formation or lags no more than a few
Gyr behind the star formation peak \citep{madau2014}. There is a smooth decline
in the values of \OmegaDust\ over the past 8 Gyr, implying that the Universe is
becoming more transparent. The physical cause of this turn-down is not
well understood; hypotheses include the net destruction of grains or their ejection
from galaxies \citep[e.g.,][]{popping2017, li2019}. Several recent works have predicted the global evolution of dust in semi-analytical models \cite[e.g.,][]{popping2017, millan-irigoyen2019, vijayan2019, graziani2019} and
by incorporating self-consistent dust physics on-the-fly into bona fide
hydrodynamic simulations \citep[e.g.,][]{bekki2015b, aoyama2018, mckinnon2018,
hou2019, li2019}. These latter works account at the sub-grid level for the processes of grain production (e.g., stellar or supernova production, ISM accretion) and destruction (e.g., by shattering / sputtering in hot gas / shocks, X-ray heating driven by AGN, astration). We show the results for dust in galaxies from two simulations in Figure
\ref{fig:omega_dust} \citep{aoyama2018,li2019}. Their outcomes in Figure
\ref{fig:omega_dust} show similar slopes characterizing the dust mass build-up
to $z \ga 2$. However, due to differing assumptions about grain production and destruction at the sub-grid level, there is significant variation between them in both the total dust density predicted and the low redshift behavior. The variations in these results are indicative of the uncertainties remaining in dust modeling in a cosmological context. The emerging field of dust measurements at high redshift will provide additional constraints on the sub-grid physics important for such modeling.

\begin{textbox}[h]
\section{Cosmic Evolution of Dust}
We have used the multi-element methods for correcting elemental depletion to estimate the amount of the metals locked into dust grains in neutral gas. We derive the dust-to-gas ratio (\dtg) in neutral absorbers, extending such measurements to lower metallicities than are available in the local Universe. We find the \hi -weighted \dtg\ increases by $\sim1$ dex from $z\approx5$ to today, in parallel to the increase in mean metallicity. We combine these \dtg\ estimates to derive the global dust density of the neutral gas over $0 \la z \la 5$, showing a gradual increase in the amount of dust in the Universe over time. Our findings are broadly in agreement with measurements of the dust density found in galaxies at $z\la2$. These measurements will provide new constraints to the next generations of hydrodynamical simulations incorporating dust physics to understand the galaxy contributors to the global build-up of dust.
\label{box:omega_dust} 
\end{textbox}

\section{LOOKING FORWARD AND CONCLUSIONS}
\label{sec:future}

In this review, we have described observational constraints on the global evolution of gas, metals, and dust over cosmic time. The progress in the field in the last two decades, both observationally and in terms of simulations, has been truly remarkable.

\subsection{Pushing from the global to the local}

The topics covered in this review refer to global quantities averaged
over the contributions from many environments. The strength (and the limitation)
of such density estimates is that they are made regardless of the
precise nature of the absorbers. However, the ultimate goal is to characterize
the connection between the components depicted in Figure \ref{fig:sketch} and
these global properties.  Important progress in this field will require breaking
down the types and properties of the environments that contribute to the metal mass
densities in Figure \ref{fig:omega_metal} as a function of redshift.

Absorbers tracing neutral gas (with \lognhi$>$20.3) are strongly associated with
galaxies. We have witnessed an explosion of the number of galaxies found to be
associated with such absorbers. While early efforts had low success rates, the
turning point in identifying and connecting galaxies to such absorbers has been
the use of 3D-spectroscopy at near-infrared ({\it VLT/SINFONI, Keck/OSIRIS}),
optical ({\it VLT/MUSE, Keck/KCWI}) and millimeter ({\it ALMA}) wavelengths
\citep[e.g.,][]{bouche2016, peroux2017, hamanowicz2020}. Among the most
surprising aspects of such work has been the detection of strong CO emission
from galaxies associated with quasar absorbers, implying large molecular masses
\citep{neeleman2016, klitsch2018}.

These galaxy identifications draw a picture in which the gas selected in
absorption is often not uniquely associated with a single luminous galaxy.
Understanding the environments that contribute to the global baryonic, metal,
and dust censuses described in this review will require the compilation of large
samples of such identifications in order to dissect the contributors
\citep{peroux2019}.  Currently, best efforts combine the dynamics,
kinematics and metallicity information to relate the neutral, ionised and
molecular phase of the gas component by component
\citep{bouche2012,rahmani2018}.

Results from simulations predicting poor metal mixing have motivated observers
to use close quasar pairs \citep{rubin2018} as well as multiple images from
gravitationally lensed background objects to probe the transverse small-scale
coherence along sightlines dozens of kpc apart \citep{chen2014, rubin2018}.
Using extended objects as background source provide continuous map of absorbers
on small scales \citep{cooke2015,lopez2018,peroux2018}.  The spatial resolution
of the instrument is a key factor for these works. In the future, an increased
number of targets will be within reach thanks to the collecting area of the next
generation of telescopes combined with IFU capabilities such as the {\it
ELT/HARMONI} instrument. At higher redshifts,
bright targets will be within reach of {\it JWST}. At lower redshifts, {\it
BlueMUSE/VLT} will provide bluer coverage with a large field-of-view thus
increasing the number of accessible targets for such detections (e.g. {\it Messier}).

Eventually, a more direct probe of the extent of gas and metals around galaxies
will come from their direct detection in emission. Circumgalactic gas has been
detected through stacking of narrow-band images of galaxies which revealed
diffuse \lya\ haloes extending up to several dozens of kpc \citep{steidel2011,
momose2016}. More recently, {\it VLT/MUSE} observations have revealed \lya\
haloes around individual emitters \citep{leclercq2017, wisotzki2018}.  The
construction of the {\it Dragonfly} experiment, able to probe low surface
brightnesses, from commercially-available cameras \citep{abraham2014}
demonstrates that new observational parameter space can be explored without
necessarily relying on novel technological development. Emission experiments,
however, constitute a considerable challenge at higher redshifts given the
cosmological surface brightness dimming scales as $(1+z)^{-4}$. Dedicated
UV-facilities may play a critical role in the characterization of metals and gas
around galaxies. An example is {\it FIREBall}, a balloon-borne 1-m telescope
coupled to an ultraviolet spectrograph \citep{hamden2019}, though larger
missions have been proposed.

Large-scale models of the $\Lambda$CDM Universe predict a structure made of sheets and filaments of dark matter arranged in a {\it cosmic web}. The gas that we have characterized in this review -- largely found within galaxy halos -- is affected by the environment of those halos \citep{kraljic2019}. The ionized gas needed to resupply the neutral and molecular gas consumed by star formation (Figure \ref{fig:omega_cond}) is ultimately drawn from the filaments within which galaxies are embedded. We can characterize the connection of this large scale structure to halos with 3D tomographic maps using absorption lines toward a sufficiently dense collection of background sources \citep{pichon2001,
lee2016}. A new generation of spectroscopic surveys will offer the prospect to reconstruct
robustly the 3D structure of the cosmic web. In particular, Subaru {\it PFS}
will map the three-dimensional gas distribution over a large cosmological volume
at $2 < z < 3$ through a large spectroscopic survey. Later, the {\it MOSAIC} instrument will benefit from the collecting area of the {\it ELT} to further expand such studies.

\subsection{Simulating Baryonic Processes in Cosmological Context}
\label{sec:simu}

From a theoretical viewpoint, the level of complexity arises from the different scales involved: from large scale cosmological environment to pc scale interstellar physics which is both physically challenging to model and time-consuming to simulate. Over the past decade, advances in numerical methodologies and computing speed have allowed extraordinary progress in our ability to simulate the formation of structures. It is a remarkable achievement that the overall physical properties (density and temperature) of the gas (specially \hi) in these models reproduce to observations of the absorbers column densities and line widths \citep[][but see \citealt{gaikwad2017}]{fumagalli2011, rahmati2014}. However, even with today's most powerful computers, we are still far from being able to self-consistently simulate the formation of molecular clouds in a cosmological context.
Simulating the multiphase interstellar medium (ISM) still remain challenging even in ``zoom-in'' simulations, so that it is necessary to make use of subgrid modules to model unresolved physical processes, such as the formation of molecular clouds, winds from dying stars, and supernovae \citep{teyssier2019}. Progress in understanding the global baryon cycle will have impact beyond understanding the evolution of star formation and galaxies. Indeed, the nature of this baryonic physics impacts the matter power spectrum and thus the inference of cosmological parameters from weak lensing measurements from next generation of surveys such as LSST, Euclid and WFIRST \citep{semboloni2011, chisari2018, foreman2019}.

By including more realistic physics, simulations will be an essential tool to interpret current observations. One of the most pressing set of questions is to understand which objects and media contribute to the global quantities presented in this review. The main contributors to \OmegaGas\ have column densities centered around \lognhi=21.0 \citep{peroux2003a,noterdaeme2012}. The question remains as to which extent these are associated with the ISM of galaxies
or with cold clumps in the CGM as well as how these contributions change with redshift. In addition, a better understanding of the cycling of baryons between \hi\ and the ionized phase of the gas will likely come from improved simulations.

The inclusion of a full treatement of CO and \Hmol\ will be crucial to interpret the flow of information on the molecular gas in the Universe afforded by ALMA and other mm facilities. Today, observations rely on a number of assumptions (Spectral Line Energy Distribution - SLED and CO-to-\Hmol\ conversion factor) to relate CO to molecular hydrogen. A better understanding of how well does CO traces \Hmol\ (as a function of temperature, density, metallicity and redshift) is imperative to make progress in this field. The most essential ingredients to be considered are non-equilibrium chemistry and cooling, both for molecular hydrogen and heavy elements, and particularly radiative transfer of ionizing and dissociating radiation \citep{richings2014b, pallottini2019}.
Together, these points will provide a better understanding of the contributors to the global evolution of molecular gas in the Universe, \OmegaMol.

The metallicity and total metal content of the highly-ionised phase of the gas continue to constitute an important observational challenge. Observers rely on tracers such as CIV, SiIV or OVI, but relating these measures to the total amount of carbon or silicon depends strongly on ionization models (and poorly-constrained assumptions about the ionization mechanism). In addition, the mixing of metal clouds could remain incomplete \citep[e.g.][]{schaye2007}. Recently, several simulations
\citep{churchill2015,peeples2019} find that in the CGM, high ionization gas seen in absorption arises in multiple, extended structures spread over $\sim100$ kpc. Due to complex velocity
fields, highly separated structures give rise to absorption at similar
velocities \citep[see also][]{bird2015}. These authors predict a mismatch
between the smoothing scales of \hi\ and high-ionisation metals. Therefore depicting a realistic distribution of metals on galaxy scale will be of paramount importance to fully understand of the cycle of baryons. 

Finally, a more complete treatment of dust production and destruction, a key catalyst for star formation, is also crucial \citep{mckinnon2018, li2019}. On global scales, it will be a salient finding to explain the observed turndown of \OmegaDust\ at low redshift. The distribution of dust about galaxies and the fraction of material expelled from bound systems will provide crucial information on our understanding of these quantities. The grain sizes and ultimately the composition of dust will further constrain the possible extinction and depletion central to the observational quantities derived in this review.

\subsection{Concluding Remarks}
\label{sec:ccl}

Probing the Universe's constituents with absorption line techniques has proven a powerful approach.  Absorption line measurements, unlike studies of emission, are uniformly sensitive at all redshifts. Modern samples of absorption lines are large and spread over the whole sky are free from effects of cosmic variance, allowing us to delineate the global evolution of the baryonic properties of the Universe. In this review, we have explored the conclusions resulting from this simple, but powerful, counting of atoms in absorption line measurements. These include constraints on the global evolution of baryons, metals, and dust over 90\% of cosmic time. Although some issues remain (see {\it Future Issues} box below), the cosmic evolution of these quantities is -- in the broadest strokes -- fairly secure (see {\it Summary Points} box below). Connecting this large scale view to the smaller scale is an exciting direction for future work.

\begin{summary}[SUMMARY POINTS]
\label{box:ccl}

\begin{enumerate}

\item This review focused on the the global quantities of baryons, metals, and dust. It sought to relate the changes in these quantities to one another, as material cycled through the various phases of baryons with cosmic time.

\item The global densities of stars and neutral gas are well-constrained to $z\approx 5$. Observations are starting to provide new constraints on the molecular hydrogen density. Though there is significant work to be done to understand the systematics associated with deriving molecular densities from current CO measurements, we have now for the first time estimates of this quantity from high-redshift until z=0.

\item The shape of the molecular density with redshift is remarkably similar to that of the
SFR density, pointing to a strong coupling between these quantities. The global density
of condensed matter in stars and neutral and molecular material has increased with time,
requiring the accretion of material from the ionized gas reservoir into these cold forms.
The baryonic accretion rate density onto dark matter halos scaled by an efficiency factor
shows a similar shape to the net condensed matter accretion rate. These results indicate
that the low-redshift decline of the star-formation history is driven by the lack of molecular
gas supply due to a drop in net accretion rate, which is itself driven by the decreased
growth of dark matter halos. These observations are in remarkable agreement with the
gas regulator model, which describes a continual cycle of baryons flowing in and out of
galaxies as a key moderator of galaxy evolution.

\item After correcting for dust depletion, the mean metallicity of the neutral gas in the Universe increases by $\approx10\times$ over the redshift range $z\approx 5$ to $\approx0$, whereas the global comoving metal density of neutral gas only increases by $\approx4.5\times$ over that same period (due to the slowly decreasing gas density).

\item The metals in neutral gas dominates the metal content of the Universe at $z\ga2.5$, consistent with containing nearly all of the metals produced by stars to that point. The significant contributors to the metal budget diversify with time, with stars and hot gas becoming increasingly important hosts of metals. Although direct observational constraints are lacking at intermediate redshifts ($1 \la z \la 2$), modern censuses indicate that the metal content of the Universe is largely accounted for.

\item Analysis of dust depletion in neutral gas provides an estimate of dust properties up to $z\approx5$. This analysis shows the dust content of neutral gas is a strong function of metallicity extending to below 1\% of the solar metallicity. The global comoving dust density of neutral gas in the Universe increases by nearly 1 dex from $z\approx5$ to $z\approx 1$ and is in agreement with estimates of dust mass in galaxies at $z < 2$.

\end{enumerate}
\end{summary}

\begin{issues}[FUTURE ISSUES]
\label{box:future}
\begin{enumerate}

\item {\bf Global evolution:} What approaches will be fruitful for improving the assessment of \OmegaGas\ at $z>3.5$? What paths will allow us to better quantify the density of {\it ionized} gas, $\Omega_{\rm H^+}$? How can we minimize the systematics in estimating \OmegaMol ? Are there better proxies for determining \OmegaMol\ (i.e., alternatives to CO)? How well can we simulate the colder phase of the gas in galaxies? How to observe directly the accretion implied by Figure \ref{fig:gas_depletion}? How to extend measurements of \OmegaDust\ (\S\ref{sec:omega_dust}) to earlier cosmic times, when the first grains are being formed? Which physical processes cause the observed evolution? Can we simulate them?

\item {\bf Missing populations:} Is our current sample missing a population of absorbers (\S\ref{sec:samplebias}) that significantly biases the characterization of \OmegaGas, \OmegaMet, and \OmegaDust ?

\item{\bf Completing the metal census:} To what redshifts do metals in stars and the hot gas in groups and clusters contribute significantly to the metal census (\S\ref{sec:omega_metal})? How do we improve on measurements of the metal content at columns below those associated with neutral gas (\S\ref{sec:missing_census})?

\item {\bf Galaxy contributors:} What galaxy populations contribute most to the baryonic densities \OmegaGas\ and \OmegaMol\ as well as to the densities \OmegaMet\ and \OmegaDust? How does the mix of contributions change with redshift?

\item {\bf Spatial distribution of metals and dust about galaxies:} How are metals and dust distributed about galaxies and beyond (Figure~\ref{fig:sketch})? How does this change with redshift? Is there pristine gas to be found? Can we reproduce the metal and dust distribution in simulations?

\end{enumerate}
\end{issues}

\section*{DISCLOSURE STATEMENT}
The authors are not aware of any affiliations, memberships, funding, or financial holdings that
might be perceived as affecting the objectivity of this review.

\section*{ACKNOWLEDGMENTS}

CP thanks the Alexander von Humboldt Foundation for the granting of a Bessel
Research Award held at MPA. JCH
recognizes support from NSF grant AST-1517353.
We are grateful to the following people for sharing data and for useful discussions:
Shohei Aoyama,
Annalisa de Cia,
Valentina D'Odorico,
Samuel Quiret,
Nicolas Lehner,
Ryan McKinnon,
Dylan Nelson,
Molly Peeples, and
Jonghwan Rhee.
We also wish to acknowledge insightful comments from
George Becker,
Nicolas Bouch\'e,
I-Da Chiang,
Stefano Ettori,
Andrea Ferrara,
Mike Fall,
Masataka Fukugita,
Michele Fumagalli,
Reinhard Genzel,
Ed Jenkins,
Andrey Kravtsov,
Varsha Kulkarni,
Bruno Leibundgut,
Piero Madau,
Brice M\'enard,
Simon Morris,
Max Pettini,
Gerg\"o Popping,
Philipp Richter,
Mike Shull,
Linda Tacconi,
Paolo Tozzi,
Juan Vladilo,
Rob Yates,
Don York,
Fabian Walter,
Simon White, and
Martin Zwaan.

%
\bibliography{araa_biblio}

\end{document}